\documentclass[aps,prb,superscriptaddress,floatfix,twocolumn]{revtex4}
\usepackage[T1]{fontenc}
\usepackage[english]{babel}
\usepackage{rotating}
\usepackage{lscape}
\usepackage{graphics}
\usepackage{graphicx}
\usepackage{dcolumn}
\usepackage{epsfig}
\usepackage{bm}
\usepackage{stmaryrd}
\usepackage{latexsym}
\usepackage{amssymb}
\usepackage{amsfonts}
\usepackage{afterpage}
\usepackage{amsmath}
\usepackage{fancybox}
\usepackage{color}
\usepackage{afterpage}
\allowdisplaybreaks
\definecolor{MyDarkBlue}{rgb}{0,0.08,0.45}

\allowdisplaybreaks
\definecolor{MyDarkBlue}{rgb}{0,0.08,0.45}

\newcommand{\su}{\uparrow}
\newcommand{\giu}{\downarrow}

\newcommand{\be}{\begin{equation}}
\newcommand{\ee}{\end{equation}}
\newcommand{\bea}{\begin{eqnarray}}
\newcommand{\eea}{\end{eqnarray}}
\newcommand{\beal}{\begin{align}}
\newcommand{\eeal}{\end{align}}
\newcommand{\ba}{\begin{eqnarray*}}
\newcommand{\ea}{\end{eqnarray*}}
\newcommand{\dagga}{{\phantom{\dagger}}}
\newcommand{\bR}{\mathbf{R}}

\newcommand{\bk}{\mathbf{k}}

\newcommand{\bRp}{\mathbf{R'}}

\newcommand{\Pj}[2]{|#1\rangle\langle #2|}

\newcommand{\fract}[2]{\frac{\displaystyle #1}{\displaystyle #2}}

\newcommand{\eqn}[1]{(\ref{#1})}

\begin{document}

\title{Strongly correlated metal interfaces in the Gutzwiller approximation}
\author{Giovanni Borghi}
\affiliation{International School for Advanced Studies (SISSA), and CRS Democritos, CNR-INFM,
Via Beirut 2-4, I-34151 Trieste, Italy} 
\author{Michele Fabrizio}
\affiliation{International School for Advanced Studies (SISSA), and CRS Democritos, CNR-INFM,
Via Beirut 2-4, I-34151 Trieste, Italy} 
\affiliation{The Abdus Salam International Centre for Theoretical Physics 
(ICTP), P.O.Box 586, I-34151 Trieste, Italy}
\author{Erio Tosatti}
\affiliation{International School for Advanced Studies (SISSA), and CRS Democritos, CNR-INFM,
Via Beirut 2-4, I-34151 Trieste, Italy} 
\affiliation{The Abdus Salam International Centre for Theoretical Physics 
(ICTP), P.O.Box 586, I-34151 Trieste, Italy}

\begin{abstract}

We study the effect of spatial inhomogeneity on the physics of a strongly correlated electron system
exhibiting a metallic phase and a Mott insulating phase, represented by the simple Hubbard model.
In three dimensions, we consider various geometries, including vacuum-metal-vacuum, a junction
between a weakly and a strongly correlated metal, and finally the double junctions metal-Mott insulator-metal and metal-strongly 
correlated metal- metal. We applied to these problems the self-consistent Gutzwiller technique 
recently developed in our group, whose approximate nature is compensated by an extreme flexibility,
ability to treat very large systems, and physical transparency. The main general result is a clear 
characterization of the position dependent metallic quasiparticle spectral weight. Its behavior  
at interfaces reveals the ubiquitous presence of exponential decays and crossovers, with decay 
lengths of clear physical significance. The decay length of metallic strength in a weakly-strongly 
correlated metal interface is due to poor screening in the strongly correlated side. The decay length 
of metallic strength from a metal into a Mott insulator (or into vacuum) is due to tunneling. In both cases, the 
decay length is a bulk property, and diverges with a critical exponent ($\sim 1/2$ in the present approximation, 
mean field in character) as the (continuous, paramagnetic) Mott transition is approached. 

\end{abstract}
\maketitle

\section{Introduction}
Metallic electron wavefunction delocalization in a lattice of atoms or molecules is 
caused by the lowering of 
electron kinetic energy and by the simultaneous 
improvement of electron-ion Coulomb attraction. By abandoning the ion cores and turning delocalized, 
an electron can in fact 
feel the potential of more than one nucleus. However, coherent electron motion is opposed by the mutual 
electron-electron Coulomb repulsion, which is higher when electrons move due to their higher chance 
of colliding when visiting the same site. 
When the first two terms prevail, the system is a conventional band 
insulator or metal, depending whether the Fermi level falls in a band gap or across one or more bands.
When the electron-electron repulsion prevails instead 
the electrons localize on their atomic or molecular sites leading to a so-called Mott insulator~\cite{Mott}. 
Despite that conceptual simplicity, properties of Mott insulators and especially of strongly correlated metals  
in the proximity of a Mott metal-insulator transition as a function of increasing correlations 
remain quite difficult to capture both theoretically 
and experimentally. Theoretically, the reason is that the Mott transition is a collective 
phenomenon, which escapes single-particle or mean field theories such as Hartree-Fock 
or density-functional-theory within the local-density approximation (LDA). Experimentally, 
additional complications such as magnetism, lattice distortions, etc., often conspire to mask the 
real nature of the Mott localization phenomenon.

Important insights into this problem have been gained in the last two decades especially thanks to dynamical mean 
field theory (DMFT).~\cite{DMFT} DMFT predicts that, as the electron-electron repulsion -- usually parametrized by a 
short-range Hubbard repulsion $U$ -- increases, the ordinary band metal evolves first to a {\sl strongly correlated metal} 
well before the Mott transition. In the strongly correlated metal
the electron spectral function undergoes a profound change, exhibiting well formed 
Mott-Hubbard side-bands coexisting with delocalized quasiparticles, the latter
narrowly centered in energy near the Fermi level. Only successively upon increasing repulsion
do the quasiparticles disappear as the Mott transition takes place at $U=U_{\rm crit}$. This intriguing prediction --  
simultaneous metallic and insulating features, exhibited on well separated energy scales --  
has stimulated a considerable experimental effort to reveal coexisting quasiparticles 
and Mott-Hubbard bands in strongly correlated metals~\cite{Sekiyama,Maiti,Mo,Sekiyama-prl,
Kamakura,Kim,Taguchi,Mo-prb,eguchi,Yano}, especially in the paradigmatic 
system V$_2$O$_3$. This is the compound where a Mott transition has been first 
discovered~\cite{Remeika} and theoretically studied~\cite{Rice,brinkman&rice}. 
At ambient temperature and pressure V$_2$O$_3$ is a correlated metal. 
It undergoes  a first-order Mott transition at $\sim T_N\simeq 155$~K to 
an antiferromagnetic insulator accompanied by a 
monoclinic distortion of the high temperature corundum structure.~\cite{Dernier} 
The paramagnetic high-temperature metal can moreover be turned into a paramagnetic Mott insulator upon 
substituting V with bulkier
Cr, (V$_{1-x}$Cr$_x$)$_2$O$_3$. For $0.005<x<0.017$ a first-order line separates the high temperature metal from the 
paramagnetic Mott insulator, which terminates with a critical point at $T\simeq 400$~K and $x\simeq 0.005$. 

Near the metal-insulator transition of (V$_{1-x}$Cr$_x$)$_2$O$_3$, 
the strongly correlated metal must of course possess well defined quasiparticles at the Fermi energy. Surprisingly, 
early photoemission experiments~\cite{Post,Smith,Shin,Zimmermann} failed to reveal the sharp quasiparticle 
peak predicted by DMFT at $E_F$. The electronic spectrum appeared instead dominated by the lower 
Mott-Hubbard band with barely a hint of metallic weight at the Fermi energy.  
It was recognized only later that photoemission in strongly correlated metals is 
highly surface-sensitive.~\cite{Sekiyama,Maiti,Kamakura,eguchi,panaccione,Sekiyama-prl,Yano} 
By increasing the photon frequency,   
which corresponds to more energetic excited photo-electrons, i.e. longer escape lengths, 
a prominent quasiparticle peak coexisting with incoherent 
Mott-Hubbard bands was eventually observed in V$_2$O$_3$~\cite{vollardo,Mo,Mo-prb}.
Quasiparticle suppression in surface-sensitive probes was 
attributed~\cite{vollardo} to surface-modified Hamiltonian parameters,
the reduced atomic coordination pushing the surface 
closer to the Mott transition than the underlying bulk. 
This conclusion, although not unreasonable, raises however a more fundamental question. 
A metal does not possess any intrinsic long-distance electronic length-scale other than the Fermi wavelength. 
Thus an imperfection like a surface can only induce a power-law decaying 
disturbance such as that associated with Friedel's oscillations. 
Since one does not expect Luttinger's theorem to break down, these oscillations should 
be controlled by the same Fermi wavelength as in the absence of interaction, 
irrespectively of the proximity of the Mott transition.  
On the other hand, a strongly correlated metal does possess an intrinsic energy scale, the parametric distance of the Hamiltonian from 
the Mott transition, and that could be associated with a length scale. 
For example, the arising of a critical length scale in association with a free energy scale
is well known in second order phase transitions. 
The surface as a perturbation may alter 
the quasiparticle properties within a depth corresponding to that characteristic length. We expect this length to be
a bulk property, the longer the closer the 
Mott transition, unlike the Fermi wavelength that remains constant. In this respect, 
it is not {\sl a priori} clear whether the recovery of bulk-quasiparticle spectral 
properties with increasing depth should be power-law, compatible with the common view of a metal as an 
inherently critical state of matter, or exponential, as one would expect by regarding the Mott transition  
as any other critical phenomena where power laws emerge only at criticality. 

Besides the interface with vacuum, which is relevant to spectroscopy, other types of interface involving correlated materials are 
attracting increasing interest. In 2004, Ohtomo and Hwang~\cite{Ohtomo} discovered that the interface between two insulating oxides, 
LaAlO$_3$ and SrTiO$_3$, is a high-mobility two-dimensional conductor that even shows superconductivity~\cite{Reyren}. 
This discovery stimulated experimental and theoretical studies on oxides heterostructures~\cite{nota-biblio}.
On the theory side, some activity has been focused either on the characterization of the electronic 
structure of these interfaces by {\sl ab-initio} LDA calculations, see e.g. Ref.~\onlinecite{pentcheva06}, as well as on 
DMFT analyses of simple models~\cite{Potthoff-1,Potthoff-2,Okamoto-Nature,Okamoto-PRB-04,Freericks,Liebsch,
Freericks-PRB-07,Rosch,Pruschke-condmat} and on combined LDA-DMFT calculations~\cite{ishida} aimed at understanding interface
correlation effects poorly described within straight LDA.   
The DMFT approaches adopted in the literature to describe this kind of situations were {\sl ad-hoc} extensions 
of the single-site DMFT~\cite{DMFT}   
to inhomogeneous systems.~\cite{Potthoff-1,Potthoff-2} In the specific example of a layered structure, 
the electron self-energy was assumed to depend, besides the frequency, also upon the layer index. 
In this scheme the self-energy is calculated by solving an auxiliary impurity model for each layer 
in which the conducting bath depends self-consistently  
on the fully-interacting impurity Green's functions not only of that given layer but also of the nearby ones. 
This additional complication with respect to conventional DMFT weighs on the numerical calculation, which is thus limited to 
few tens of layers. Although this is adequate for
the interface between two insulators, such as that studied by Ohtomo and Hwang~\cite{Ohtomo}, it is generally 
insufficient in other cases, such as the 
surface effects in the interior of a correlated metal,~\cite{Marsi}  
or any other interface involving at least one metal. 

Recently, we 
proposed an alternative theoretical approach to interface problems,~\cite{Ours} 
based on the extension of the Gutzwiller wavefunction and approximation~\cite{Gutzwiller1,Gutzwiller2} to inhomogeneous situations. 
The method, although 
a further approximation beyond DMFT, hence in principle less accurate, is much more agile, and 
can treat without effort hundreds of layers. Thus it can be used as a complementary tool to extrapolate DMFT results to large sizes, 
otherwise unaccessible by straight DMFT. 

In this work, we shall extend the analysis of Ref.~\onlinecite{Ours} for the vacuum/correlated-metal interface 
to other model interfaces that might be relevant for experiments: the junction between two different correlated metals and 
the tunneling between two metallic leads through a strongly correlated, possibly Mott insulating, region. 
Although both cases 
were in fact previously studied by DMFT~\cite{Rosch,Freericks,Pruschke-condmat}, 
the results were interpreted in contrasting ways. While Helmes {\sl et al.}~\cite{Rosch} concluded that the Mott insulator is 
impenetrable to the electrons coming from the metallic leads, Zenia {\sl et al.}~\cite{Pruschke-condmat} drew the opposite conclusion 
that a conducting channel always open up inside the insulator at sufficiently low temperature.    
The present study, which is certainly less accurate than DMFT but can deal with much larger 
sizes, will also serve to clarify this issue. 
In particular, the large sizes allow us to address the asymptotic 
behavior and to identify the magnitude and interface role of the critical length associated with the 
bulk Mott transition.

The paper is organized as follows. In section~\ref{method} we introduce the model Hamiltonian, which is a Hubbard model 
with layer dependent parameters, and a Gutzwiller variational scheme adapted for such an inhomogeneous situation. 
We then study in section~\ref{results} three different slab geometries: (a) strong correlated metal--vacuum interface; 
(b) junction between two different correlated metals; (c) a Mott insulator or a strongly correlated metal sandwiched between two 
weakly correlated metals.   
In the first two cases we find that the perturbation induced by the surface inside the bulk of the correlated metal 
decay exponentially at long distances. The length scale $\xi$ that controls this decay is a bulk property that depends in our 
simplified model only on $U_{\rm crit}-U$ and diverges on approaching the Mott transition like  
$\xi\sim \left(U_{\rm crit}-U\right)^{-\nu}$, with a mean-field like exponent $\nu\simeq 0.5$. The last case (c) is more interesting.
Either when the central region, of width $d$, is a strongly correlated metal, $U_{\rm center}<U_{\rm crit}$ or when it is a Mott 
insulator, $U_{\rm center}>U_{\rm crit}$, the effects of the two metal leads are found to decay exponentially over a length $\xi$.  
Just like in cases (a) and (b) above, $\xi$ is only controlled 
by the distance from Mott criticality, i.e. 
\[ 
\xi \sim \left|U_{\rm crit}-U_{\rm center}\right|^{-0.5},
\]
which therefore appears naturally as a correlation length that is finite on both sides of the transition. 
However, while the quasiparticle weight saturates to a finite constant determined by $U_{\rm center}<U_{\rm crit}$ 
and independent of $d$   
when the central region is a strongly correlated metal, in the opposite case of a Mott insulator the quasiparticle weight 
saturates to a finite value exponentially small in $d$. Interestingly, right at criticality, $U_{\rm center}=U_{\rm crit}$, 
the saturation value decays power law in $d$. Finally, section~\ref{Conclusions} is devoted to concluding remarks. 
For a better understanding of our numerical data, a simple 
analytical model for the spatial dependence of quasiparticle weight is set up in appendix~\ref{analytical}, while  
in appendix~\ref{friedel} we discuss the effects of electron-electron interaction on the physics 
of Friedel's oscillations near surfaces and junctions within the Gutzwiller approximation. 

\section{Model and method}\label{method}

In order to address the generic interface features of a 
a strongly correlated metal, we consider the simplest Hamiltonian exhibiting a Mott transition, 
namely the Hubbard model 
\bea
H &=& -\sum_{<\bR\bRp>\sigma}\, t_{\bR\bRp}\,
\Big(c^\dagger_{\bR\sigma}c^\dagga_{\bRp\sigma} + H.c.\Big) \nonumber\\ 
&& +  \sum_\bR\,\epsilon_\bR n_\bR + U_\bR\,n_{\bR\su}n_{\bR\giu},
\label{Ham}
\eea
where $<\bR\bRp>$ denotes nearest neighbor sites, $c^\dagger_{\bR\sigma}$ and $c^\dagga_{\bR\sigma}$ 
creates and annihilates, respectively, an electron at site $\bR$ with spin $\sigma$, and finally  
$n_{\bR\sigma}=c^\dagger_{\bR\sigma}c^\dagga_{\bR\sigma}$ and $n_\bR = n_{\bR\su} + n_{\bR\giu}$. 
In our inhomogeneous system, all Hamiltonian parameters are allowed to be site dependent. 
For interfaces, we shall assume an $N$-layer 
slab geometry where all parameters are constant 
within each layer, identified by a layer coordinate $z=1,\dots,N$ but generally different from layer to layer. For instance, the 
hopping between nearest neighbor sites $\bR$ and $\bRp$ within layer $z$ depends only on $z$, i.e. $t_{\bR\bRp} = t(z)$, 
while if $\bR$ and $\bRp$ belong to nearby layers, e.g. $z$ and $z\pm 1$, then $t_{\bR\bRp} = t(z,z\pm 1)=t(z\pm 1,z)$.  

We study the Hubbard Hamiltonian \eqn{Ham} in the non-magnetic (also called paramagnetic) sector by means of a Gutzwiller 
type variational wavefunction 
\be
|\Psi\rangle = \prod_\bR\,\mathcal{P}_\bR\, |\Psi_0\rangle,\label{GWF}
\ee
where $|\Psi_0\rangle$ is a paramagnetic Slater determinant. Because 
of our choice of layer-dependent parameters, the operator $\mathcal{P}_\bR$ has the general expression
\be
\mathcal{P}_\bR = \sum_{n=0}^2\, \lambda_n(z)\,\Pj{n,\bR}{n,\bR},
\label{P_R}
\ee
where $\Pj{n,\bR}{n,\bR}$ is the projector at site $\bR=(x,y,z)$, ($x$ and $y$ are intralayer coordinates), onto configurations 
with $n$ electrons (note that $\Pj{1,\bR}{1,\bR} \equiv \sum_\sigma c^\dagger_{\bR\sigma}
\Pj{0,\bR}{0,\bR}c^\dagga_{\bR\sigma}$), and $\lambda_n(z)$ are layer-dependent variational parameters.  
We calculate quantum averages 
on $|\Psi\rangle$ using the so-called Gutzwiller approximation~\cite{Gutzwiller1,Gutzwiller2}, 
(for details see e.g. Ref.~\cite{mio} whose notations we use hereafter) and require that  
\bea
\langle \Psi_0|\mathcal{P}^2_\bR|\Psi_0\rangle &=& 1,\label{1}\\
\langle \Psi_0|\mathcal{P}^2_\bR\,n_{\bR\sigma}|\Psi_0\rangle &=& \langle \Psi_0|n_{\bR\sigma}|\Psi_0\rangle
\equiv \frac{n(z)}{2}.\label{2}
\eea 
Explicitly, these two conditions imply that
\bea
1 &=& \left(1-\frac{n(z)}{2}\right)^2\lambda_0(z)^2 \nonumber \\
&& + n(z)\left(1-\frac{n(z)}{2}\right)\lambda_1(z)^2 + \frac{n(z)^2}{4}\lambda_2(z)^2,\label{1-bis}\\
n(z) &=& n(z)\left(1-\frac{n(z)}{2}\right)\lambda_1(z)^2 
+ 2\frac{n(z)^2}{4}\lambda_2(z)^2.\label{2-bis}
\eea
We note that $n(z)$ is fixed once the uncorrelated variational wavefunction $|\Psi_0\rangle$ is given. In reality we 
find more convenient to treat $n(z)$ as an additional variational parameter, and constrain $|\Psi_0\rangle$ to span all 
paramagnetic Slater determinants that have a fixed local charge density $n(z)$. 
The average value of \eqn{Ham} within the Gutzwiller approximation is accordingly given by~\cite{mio,Bunemann}
\bea
E&=&\fract{\langle \Psi|\,H\,|\Psi\rangle}{\langle \Psi|\Psi\rangle} \simeq \sum_\bR\, U_\bR \,
\frac{n(z)^2}{4}\,\lambda_2(z)^2 + \epsilon_\bR\,n(z)\label{E-var}\\
&& -\!\!\sum_{<\bR\bRp>\sigma}t_{\bR\bRp}\,R(z)\,R(z')\,
\langle \Psi_0 |c^\dagger_{\bR\sigma}c^\dagga_{\bRp\sigma} + H.c.|\Psi_0\rangle,\nonumber 
\eea
where 
\be
R(z) = \left(1-\frac{n(z)}{2}\right)\,\lambda_0(z)\lambda_1(z) + \frac{n(z)}{2}\,\lambda_1(z)\lambda_2(z),\label{R(z)-exp}
\ee
plays the role of a wavefunction renormalization factor, whose square can be regarded as the actual layer-dependent 
quasiparticle weight, $Z(z)=R^2(z)$. Because of Eqs.~\eqn{1-bis}, \eqn{2-bis} and \eqn{R(z)-exp}, one can express 
\[
\lambda_n(z) = \lambda_n\left[R(z),n(z)\right],
\]
as functional of the two variational functions $R(z)$ and $n(z)$. 
Furthermore, the single-particle wavefunctions that 
define the Slater determinant $|\Psi_0\rangle$ can be chosen, for a slab geometry, to have the general expression 
\[
\phi_{\epsilon\bk_{||}}(\bR) = \sqrt{\frac{1}{A}}\,\mathrm{e}^{i\bk_{||}\cdot\bR}\, \phi_{\epsilon\bk_{||}}(z),
\]
where $A$ is the number of sites per layer and $\bk_{||}$ the momentum in the $x$-$y$ plane. The minimum of $E$, Eq.~\eqn{E-var}, 
can then be obtained by searching for saddle points with respect to the variational parameters $R(z)$, $n(z)$  
and $\phi_{\epsilon\bk_{||}}(z)$, the latter subject to the constraint 
\[
\frac{2}{A}\sum^{occupied}\, \left|\phi_{\epsilon\bk_{||}}(z)\right|^2 = n(z),
\]
the sum running over all occupied states in the Slater determinant. 

Considerable simplifications arise if we further assume a bipartite lattice with a Hamiltonian \eqn{Ham} invariant under the 
particle-hole transformation
\[
c^\dagga_{\bR\sigma}\rightarrow \sigma\,(-1)^R\,c^\dagger_{\bR -\sigma},
\]
where $(-1)^R$ is $+1$ on one sublattice and $-1$ on the other. This symmetry requires $\epsilon_\bR = 0$ in \eqn{Ham} and implies 
$n(z)=1$ hence $\lambda_0(z) = \lambda_2(z)$ and $\lambda_1(z)^2 = 2 - \lambda_0(z)^2$. 
In this case the saddle point is simply obtained by solving the coupled equations
\begin{widetext}
\bea
&&\epsilon\,\phi_{\epsilon\bk_{||}}(z) = R(z)^2\,\epsilon_{\bk_{||}}(z)\, \phi_{\epsilon\bk_{||}}(z) 
- R(z)\,\sum_{p=\pm 1}\, t(z,z+p)\,R(z+p\,a)\,\phi_{\epsilon\bk_{||}}(z+p\,a), \label{uno}\\
&&R(z) = \fract{4\sqrt{1-R(z)^2}}{U(z)A}\,\sum_{\epsilon\, \bk_{||}}^{occupied}\Bigg[-2R(z)
\epsilon_{\bk_{||}}(z)\phi_{\epsilon\bk_{||}}(z)^2 
+\phi_{\epsilon\bk_{||}}(z)\sum_{p=\pm a} t(z,z+p)R(z+p\,a)\phi_{\epsilon\bk_{||}}(z+p\,a)\Bigg],\label{due}
\eea
\end{widetext}
where $\epsilon_{\bk_{||}}(z)=-2t(z)\,\left(\cos k_x a + \cos k_y a\right)$. The first equation has
the form of a Schr{\oe}dinger equation which the single-particle wavefunctions $\phi_{\epsilon\bk_{||}}(z)$ 
must satisfy, the quasiparticle hopping now depending parametrically on $R(z)$.  The second equation has been 
intentionally cast in the form of a map $R_{j+1}(z)=F\left[R_j(z),R_j(z+a),R_j(z-a)\right]$ 
whose fixed point we have verified to coincide with the actual solution of 
\eqn{due} in the parameter region of interest. 

In spite of the various assumptions above, solving this saddle point problem remains in principle formidable. 
Fortunately, Eqs.~\eqn{uno} and \eqn{due} can in fact be solved relatively easily, by the following iterative procedure. 
First solve the Schr{\oe}dinger equation at fixed $R_j(z)$; next find the 
new $R_{j+1}(z)$ using the old $R_j(z)$ and the newly determined 
wavefunctions $\phi_{\epsilon\bk_{||}}(z)$. With the new $R_{j+1}(z)$, 
repeat the above steps and iterate until some desired level of convergence is reached.   
Because of the large number of variational parameters, this iterative scheme 
is much more efficient than -- but fully equivalent to -- a direct minimization 
of $E$, Eq.~\eqn{E-var}. Away from particle-hole symmetry, the saddle point equations get more 
involved but the solution can be obtained along the same lines. 

Before concluding, we recall for future use the Gutzwiller approximation results for the 
Mott transition at particle-hole 
symmetry in the homogeneous case, $\epsilon_\bR=0$, $t_{\bR\bRp}=t$ and $U_\bR=U$, i.e. when the variational 
parameters $\lambda_n(z)$ are $z$-independent.  
In this case, the solution of Eqs.~(\ref{uno}) and (\ref{due}) is trivial. The critical values $U=U_{\rm crit}$ at the 
Mott transition are $U_{\rm crit}=32 t/\pi$ (for a linear chain), $U_{\rm crit}=128 t/\pi^2$ (for a square lattice),
$U_{\rm crit}=16 t$ (for a cubic lattice). 
The quasiparticle weight $Z$ in terms of the electron-electron interaction $U$ has the simple expression 
\begin{equation}\label{Z_of_U}
Z=R^2=1-\frac{U^2}{U^2_{\rm crit}}\,,
\end{equation}
linearly vanishing at the Mott transition.~\cite{brinkman&rice}

\section{Interfaces in the 3D Hubbard model: results}\label{results}

We use the technique just exposed to study 3D simple cubic Hubbard model interfaces in a slab geometry 
with in-plane ($xy$) translational symmetry and layer($z$)-dependent Hamiltonian parameters.  
We assume for simplicity particle-hole symmetry and site-independent 
hoppings $t_{\bR\bRp}=t$ throughout, so that the only source of inhomogeneity is a layer-dependent $U(z)$. 
Therefore the minimization procedure amounts to solve the coupled equations \eqn{uno} and \eqn{due} with constant hoppings.   
Technically, we diagonalized the in-plane $k$-dependent Hamiltonian \eqn{uno} at
every point of a Monkhorst-Pack $k$-grid~\cite{monkhorst_pack}. The two-dimensional grid used was $32\times 32$,  
chosen so as to yield well converged values not just for the quasiparticle weight (for which a $4\times 4$ grid was sufficient) 
but also for the hopping matrix element for the geometries and interaction parameters considered.
At every iteration $j$, we choose for the convergence indicator
\begin{equation}
Q_j = \frac{1}{N}\left(\sum_{i=0}^{N} \left| Z_{j}(i)-Z_{j-1}(i)\right|\right)\,
\end{equation}
a threshold of $10^{-6}$. This corresponds to a relative energy convergence of less than $10^{-7}$.
The calculations of the spatial dependence of the hopping matrix elements  (see appendix~\ref{friedel} ) 
were instead performed with a denser $k$-grid of $64\times 64$ $k$-points.

\begin{figure}[t]
\includegraphics[width=8.1cm]{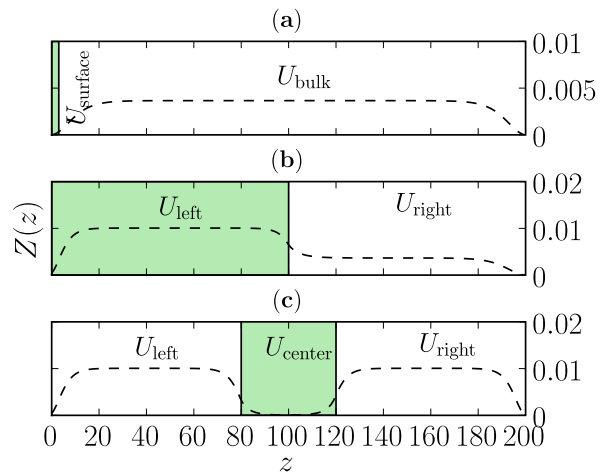}
\caption{(Color online) The three different inhomogeneities studied in this paper: (a) free surface geometry, 
(b) junction between metals with different strength of correlation, (c) Mott (or strongly correlated metallic)  
slab sandwiched between metallic leads (sandwich geometry). The values for $U$ in all the three cases shown are: 
(a) $U_{\rm surface} = 20 t$, $U_{\rm bulk} = 15.9712 t$; 
(b) $U_{\rm left}=15.9198t$, $U_{\rm right}= 15.9712 t$; 
(c) $U_{\rm left}=U_{\rm right} = 15.9198t$, $U_{\rm center}=16.0288t$ (which is the case of a Mott central slab). 
In panel (c) the region with electron-electron interaction $U=U_{\rm center}$ is indicated by the green-shaded area. 
}\label{Fig_geometries}
\end{figure}

We consider the three different geometries displayed in Fig.~\ref{Fig_geometries}: 

\begin{enumerate}
\item[(a)] {\sl Correlated metal-vacuum interface}: a correlated metal ($U_{\rm bulk}<U_{\rm crit}$, where 
$U_{\rm crit}=16t$ is the critical value of $U$ at the Mott transition in the cubic lattice) 
with a stronger correlated surface ($U_{\rm surface}>U_{\rm crit}$).
\item[(b)] {\sl Weakly correlated metal-strongly correlated metal interface}: a junction between a moderately 
correlated metal ($U_{\rm left}<U_{\rm crit}$) and a strongly correlated metal
($U_{\rm right} \lesssim U_{\rm crit}$).
\item[(c)] {\sl Metal-Mott insulator-metal 
double junction}: a Mott insulator $U_{\rm center} \gtrapprox U_{\rm crit}$ or a strongly correlated metal 
$U_{\rm center} \lesssim U_{\rm crit}$ sandwiched between two moderately 
correlated metallic leads $U_{\rm left}=U_{\rm right}<U_{\rm crit}$.
\end{enumerate}

The dashed lines in the panels of Fig.~\ref{Fig_geometries} show the quasiparticle weight $Z(z)$ calculated for a $N=200$ layer slab 
in the three geometries with the Hamiltonian parameters: 
\begin{itemize}
\item[{\it panel (a)}] $U_{\rm bulk} = U(z>1)=15.9712t$ and $U_{\rm surface}=U(z=1)= 20t$. The bulk is a strongly correlated
metal very close to the Mott transition, the right surface has the same $U$ as the bulk while the left surface a higher value well 
inside the Mott insulating range.
\item[{\it panel (b)}] $U_{\rm left}=U(z\leq 100)= 15.9198t$ and $U_{\rm right}=U(z>100)=15.9712t$; The left metal is much less
correlated than the right metal.
\item[{\it panel (c)}] $U_{\rm right}=U(z\leq 80)=U_{\rm left}=U(z>120)=15.9198t$ and $U_{\rm center} = U(80<z\leq 120)=16.0288t$.
Left and right leads are moderately correlated metals, the central region is Mott insulating.
\end{itemize} 

We now discuss each case separately.

\subsection{Geometry (a): Correlated metal-vacuum interface}\label{firstcase}

\begin{figure}
\includegraphics[width=8cm]{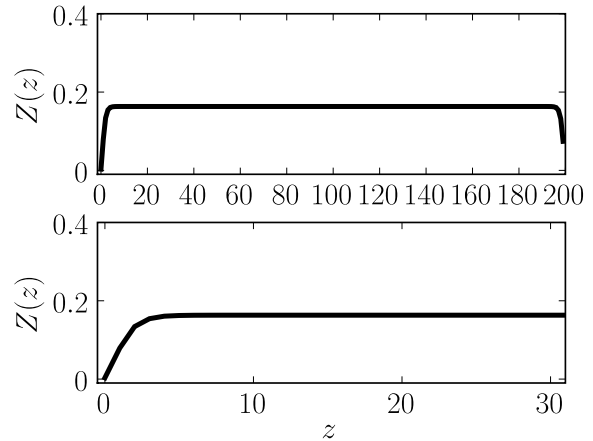}
\caption{Spatial dependence of $Z(z)$  for $U_{\rm surf}= 20t$ at $z=0$ 
and $U_{\rm bulk}= 14.6642t$, for any $z>0$. 
The lower panel is the same as the upper one zoomed close to the surface.\label{Fig_surf_1}}
\end{figure}
\begin{figure}
\includegraphics[width=8cm]{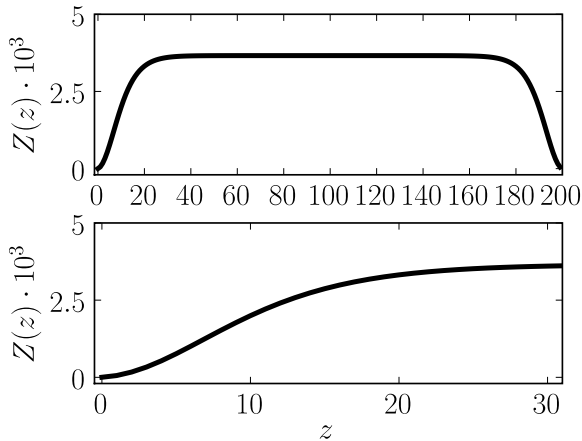}
\caption{Same as Fig.~\ref{Fig_surf_1}, for $U_{\rm surf}= 20t$ and $U_{\rm bulk}= 15.9712t$.}\label{Fig_surf_2}
\end{figure}

This is the simple surface case, 
$U(z>1) = U_{\rm bulk}<U_{\rm crit}$ and 
$U(z=1) = U_{\rm surface} > U_{\rm crit}$, previously studied in Ref.~\onlinecite{Ours}. 
Looking at Figs.~\ref{Fig_surf_1} and~\ref{Fig_surf_2}, with  
values of $U_{\rm surf}=20t$, and $U_{\rm bulk}=9.6t$ and $U_{\rm bulk}=15.97118t$, respectively,  
we observe that: 
\begin{itemize}

\item[{\sl i)}] The value of $Z(z)$ at the center of the slab, close to the bulk value, decreases monotonically to zero
while $U_{\rm bulk}$ approaches $U_{\rm crit}$. 
Due to the finite slab thickness $N$, the actual value of $U$ at which 
$Z(z)$ vanishes everywhere is slightly smaller than 
the bulk value $U_{\rm crit}=16 t$ for an infinite system, but tends to it as $N$ increases. In this limit, 
the dependence of $Z_{\rm bulk}=Z(z=N/2)$ upon $U_{\rm bulk}$ is described by Eq.\eqn{Z_of_U}.

\item[{\sl ii)}] $Z(z)$ decreases dramatically while approaching the surfaces, 
both the extra-correlated left surface $z=1$, and the regular bulk-like one at 
$z=N$. In fact, within the Gutzwiller approximation, the effective 
interaction strength at a given site is the value of $U$ relative to the average hopping energy at that 
site. The reduced surface coordination lowers the overall hopping energy of a surface site, and 
hence effectively strengthens the 
surface interaction. 
The same effect would be obtained by decreasing the hopping at the surface. We note however that, so long as
$Z$ remains finite in the interior of the slab, $Z$ remains 
finite, even if very small, also at the surface: there cannot be truly insulating surfaces coexisting with a metallic bulk. 
The reason is that, if we assume initially such an insulating surface,
then simple tunneling from the underlying bulk will bring the metallic quasiparticle weight to
a nonzero value, however small.
   
\item[{\sl iii)}] The steep decay of $Z(z)$ 
at the surfaces at $z=1$ and $z=N$ gets more and more gradual as $U_{\rm bulk}\to U_{\rm crit}$.
\end{itemize}

As found in Ref.~\onlinecite{Ours}, the behavior of $R(z)=\sqrt{Z(z)}$ can be well described 
by an exponential 
\be
R(z) = R_{\rm bulk} +\left(R_{\rm surf}-R_{\rm bulk}\right)\,e^{-(z-1)/\xi},
\label{R(z)-expo}
\ee 
\begin{figure}
\includegraphics[width=8cm]{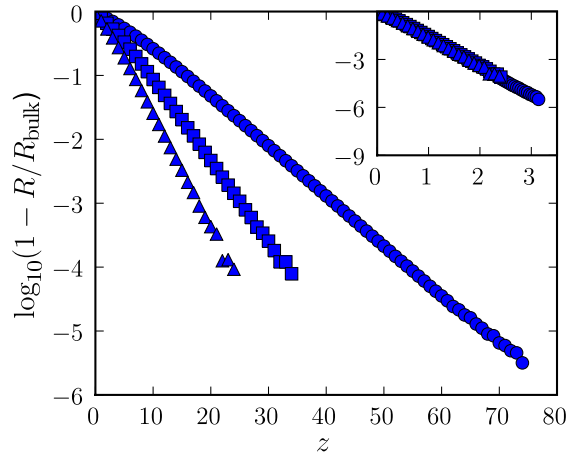}
\caption{(Color online) Plot of $\log(1-R/R_{\rm bulk})$ versus $z$ for $U=15.97118t$ (circles), 
$U=15.9198t$ (squares), $U=15.84242t$ (triangles). In the inset the same data are plotted with 
respect to $z\,(1-U/U_{\rm crit})^{0.5}$.}\label{Fig_proof_exp}
\end{figure}
where $R_{\rm bulk} = R(z=N/2)$ and $R_{\rm surf}<R_{\rm bulk}$. In Appendix A we 
actually derive a more involved analytical expression for $R(z)$ that fits well the numerical data, see Eq.~\eqn{R(z)-u<1}. 
The surface value, $R_{\rm surf}$, and the surface metallic quasiparticle weight $Z_{\rm surf}=R_{\rm surf}^2$, 
are much smaller than the bulk ones but, as previously mentioned, they can 
vanish only when $R_{\rm bulk}$ becomes strictly zero, 
for $U_{\rm bulk} > U_{\rm crit}$. 
For any $U_{\rm bulk}<U_{\rm crit}$, there is a  
surface {\sl dead layer}~\cite{Ours}, which is much less metallic than the bulk, whose thickness $\xi(U)$ depends only on bulk properties, 
and diverges for $U_{\rm bulk}\to U_{\rm crit}$ in the critical form 
\be
\xi \sim \left(U_{\rm crit}-U_{\rm bulk}\right)^{-\nu}.\label{eta}
\ee
Therefore $\xi$ may be identified with the correlation length characteristic of the bulk Mott transition. 
Numerically, we find $\nu=0.53\pm 0.3\simeq 0.5$, a typical mean field critical exponent compatible with the simple Gutzwiller 
approximation.
In Fig.~\ref{Fig_proof_exp} we plot the logarithm of the difference between $R$ and $R_{\rm bulk}$, which clearly shows the 
exponential decay for three values of $U$. In the inset of the same figure we plot the same quantity as function 
of a rescaled coordinate $z\to z\,(1-U/U_{\rm crit})^{\nu}$ with $\nu=0.5$: all data fall on the same curve thus 
substantiating our statement on the $U$-dependence of the correlation length.
Our finding of an exponential recovery of the quasiparticle weight inside the bulk in place of the expected 
Friedel-like power-law behavior offers a unique opportunity to experimentally access the critical 
properties of the Mott transition. 
Photoemission experiments~\cite{Marsi} show that the surface depletion of metallic electron spectral weight 
in V$_2$O$_3$ propagates inside the interior of the sample for an anomalously  
large depth of many tens of Angstrom beneath the surface,  in qualitative agreement  
with our results. Further experiments would be desirable to follow the behavior 
of this length scale upon approaching this and other Mott transitions and verify our prediction. 

We end by noting that the calculated $Z(z)$ shows an upward curvature near the surface ($z=0$), see Fig.~\ref{Fig_surf_2} 
and also Eq.~\eqn{Z(z)-small-z} in the appendix. This is 
unlike earlier results obtained by the so-called linearized DMFT~\cite{Potthoff-2}, displaying instead a linear growth of 
$Z(z)$ near the surface and very close to criticality. 
Besides a qualitative agreement with the upward curvature observed in 
photoemission,~\cite{Marsi} which could  
be coincidental since the real V$_2$O$_3$ is much more 
complicated than our simple one-band Hubbard model, we do not see strong arguments of principle supporting 
either approaches. Both Gutzwiller and linearized DMFT are based on rather uncontrolled approximations. More reliable 
techniques, such as straight DMFT or Quantum Monte Carlo calculations on large size systems,  
would be needed to clarify this aspect; 
but this is perhaps not important enough. What is more important is that, just like our approach, 
also linearized DMFT yields, as we checked, to a length controlling  
the depth of the surface perturbed region that diverges at the Mott transition.

\subsection{Geometry (b): Weakly correlated metal-strongly correlated metal interface}\label{secondcase}

\begin{figure}
\includegraphics[width=8cm]{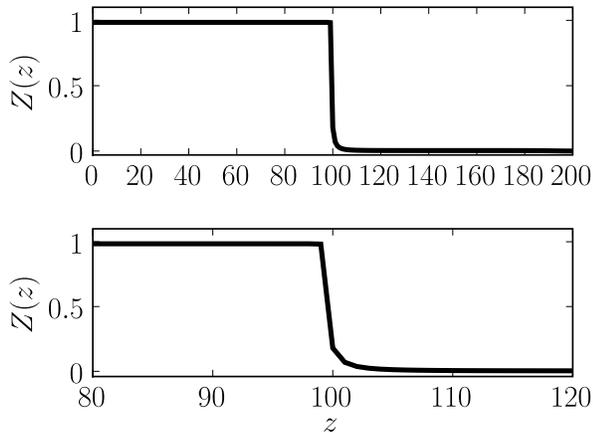}
\caption{Spatial dependence of $Z(z)$ for $U_{\rm left}= 2t$ and $U_{\rm right}= 15.9712t$. 
The lower panel shows the same data as the upper one but  
closer to the interface.\label{Fig_junction_1}}
\end{figure}
\begin{figure}
\includegraphics[width=8cm]{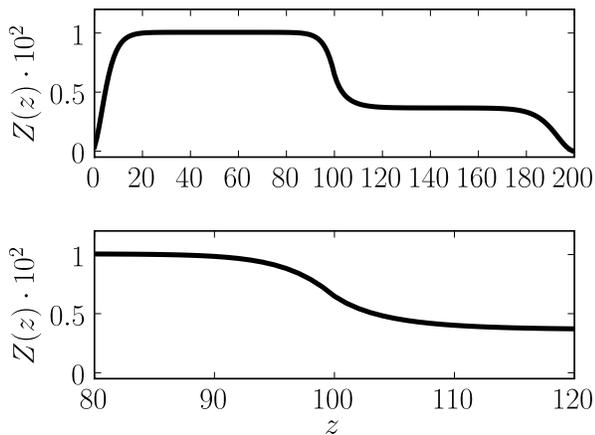}
\caption{Same as in Fig.~\ref{Fig_junction_1}, for $U_{\rm left}= 15.9198t$ and $U_{\rm right}= 15.9712t$.}\label{Fig_junction_2}
\end{figure}

The junction between a metal and a Mott insulator or a strongly correlated metal was studied recently 
by Helmes, Costi and Rosch~\cite{Rosch}, who used the numerical renormalization group as DMFT impurity solver. 
With our simpler method we can address a broader class of interfaces, including 
the general case of a correlated metal-correlated metal junction, with different values of electron-electron 
interaction in the left ($U_{\rm left}$) and right ($U_{\rm right}$) leads. 
The system we consider, see Fig.~\ref{Fig_geometries}(b), is made of two blocks $100$ layers each, 
and the junction center is at $z=N/2$. Figs.~\ref{Fig_junction_1} and~\ref{Fig_junction_2} show the $z$ dependence 
of the quasiparticle weight for fixed $U_{\rm right}\simeq U_{\rm crit}$ and two different values of $U_{\rm left}<U_{\rm right}$. 
Even if $U(z)$ is changed stepwise from left to right, we find that the closer $U_{\rm left}$ is to $U_{\rm crit}$, 
the smoother the function $Z(z)$ for $z<N/2$. 
On the right side of the junction, after a characteristic length $\xi_{\rm right}$, the quasiparticle weight $Z$ reaches 
exponentially its bulk value. 
We find for $R(z>N/2)$ a layer dependence well represented by the form (for a better fit see Eq.~\eqn{R(z)-u<1} with the 
minus sign) 
\begin{equation}
R(z) = R_{\rm right} + \left(R_{\rm left}-R_{\rm right}\right) \mathrm{e}^{-(z-N/2)/\xi_{right}}.
\end{equation}
The dependence of $\xi_{\rm right}$ on $U_{\rm right}$ is again given by Eq.~\eqn{eta}, i.e 
$\xi_{\rm right} \propto \left(U_{\rm crit}-U_{\rm right}\right)^{-\nu}$ ($\nu\approx 0.5$). By symmetry, the same holds  
in the left side too, upon interchanging the subscripts right and left.

Our results for weak $U_{\rm left}$ and $U_{\rm right}\alt U_{\rm crit}$ can be directly compared with those of  
Helmes {\sl et al.}~\cite{Rosch}, who proposed that a strongly correlated slab, our right lead with 
$U_{\rm right}\simeq U_{\rm crit}$, in contact with a non interacting metal, our left lead, has a quasiparticle 
weight $Z(x)$ that, close to criticality, has a scaling behavior   
\be
x^2\, Z(x) \simeq C\, f\left(x\left|\fract{U-U_{\rm crit}}{U_{\rm crit}}\right|^{1/2}\right),
\label{Z-Rosch}
\ee
where $f(0)=1$ and $x$ is the distance from the interface, translated in our notation $x=z-N/2$ and $U=U_{\rm right}$.  
The prefactor $C\simeq 0.008$ and the asymptotic behavior $f(\zeta\to\infty)=0.15 \zeta^2$ of the scaling function 
were extracted by a DMFT calculation with a 40 layer correlated slab in contact with a 20 layer almost uncorrelated metal~\cite{Rosch}. 

\begin{figure}
\includegraphics[width=8cm]{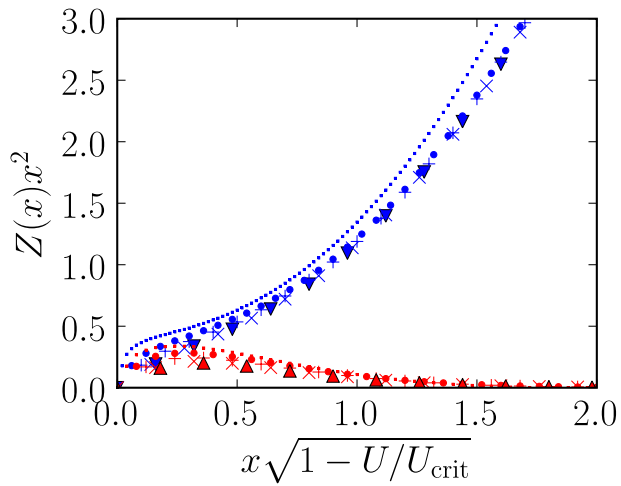}
\caption{(Color online) Plot of $Z(x)\,x^2$ versus the renormalized coordinate $x\,\sqrt{|1-U/U_{\rm crit}|}$ 
for $U<U_{\rm crit}$ (upper blue curves: $U=15.7939t$ triangles, $U=15.8424t$ crosses, $U=15.9198t$ pluses, 
$U=15.9712t$ points, $U=15.9968$ tiny dots) and $U>U_{\rm crit}$ (lower blue curves: $U=16.2571t$ triangles, 
$U=16.2035t$ crosses, $U=16.1148t$ pluses, $U=16.0511t$ points, $U=16.0128$ tiny dots). 
This figure can be compared with the inset of Fig.~3 in reference~\onlinecite{Rosch} }\label{Fig_scaling_rosch}
\end{figure}

We show in Fig.~\ref{Fig_scaling_rosch} the quantity $x^2\,Z(x)$ extracted by our Gutzwiller technique and plotted versus 
$x\left|1-U/U_{\rm crit}\right|^{1/2}$ for different $U$'s across the Mott transition value. The results are qualitatively 
similar to those of Ref.~\onlinecite{Rosch}, but differs in two aspects. First of all we find that 
$f(\zeta)$ defined in Eq.~\eqn{Z-Rosch} shows a plateau only when 
\[
z_* \ll x \ll \left|1-\fract{U}{U_{\rm crit}}\right|^{-1/2},
\]
where an approximate expression for the offset value $z_*$ is given in the appendix~\ref{analytical_int_metallic}, see 
Eqs.~\eqn{zeta-u<1} and \eqn{zeta-u>1}. For $x\ll z_*$, $f(\zeta)\sim \zeta^2$ so that $Z(x)$ approaches its surface value 
at the interface. In our data the crossover between the two different regimes is clearly visible, unlike in Ref.~\onlinecite{Rosch}. 
More seriously, the coefficient $C\simeq 0.08$ found by Helmes {\sl et al.}~\cite{Rosch} is almost two orders of magnitude 
smaller than our, which is numerically around $\simeq 0.4$. [The approximate analytical expression discussed 
in the appendix~\ref{analytical} give a slightly larger value of $2/3$, see \eqn{intercept-u<1} and \eqn{intercept-u>1}].        
In the same appendix we also show that, within the linearized DMFT approach introduced by Potthoff and Nolting~\cite{Potthoff-2} 
one would extract yet another value of the coefficient $C=9/11\sim 0.82$, of the same order as ours, and again larger than 
that found by Helmes {\sl et al.}~\cite{Rosch}. This disagreement is not just quantitative. 
Mainly because of the smallness of the prefactor, Helmes and coworkers~\cite{Rosch} concluded that the strongly correlated slab 
with $U\simeq U_{\rm crit}$ hence $Z_{\rm bulk}=Z(x\to\infty)\ll 1$ is very weakly affected by the proximity of the 
good metal, a conclusion later questioned by Zenia {\sl et al.}~\cite{Pruschke-condmat}, who however 
considered a different geometry. Our results, as well as those that could be obtained by linearized DMFT, 
do not allow any such drastic conclusion. Yet, since straight DMFT should be more reliable 
than either linearized DMFT or our Gutzwiller approach, it is likely that our $Z(x)$ is strongly overestimated 
and that  Helmes {\sl et al.}'s conclusions are basically correct. It seems worth investigating further this important question 
with full DMFT on wider slabs.    

\subsection{Geometry (c): Correlated metal-Mott insulator (Strongly correlated metal)-correlated metal double 
junction}\label{thirdcase}

\begin{figure}[t]
\includegraphics[width=8cm]{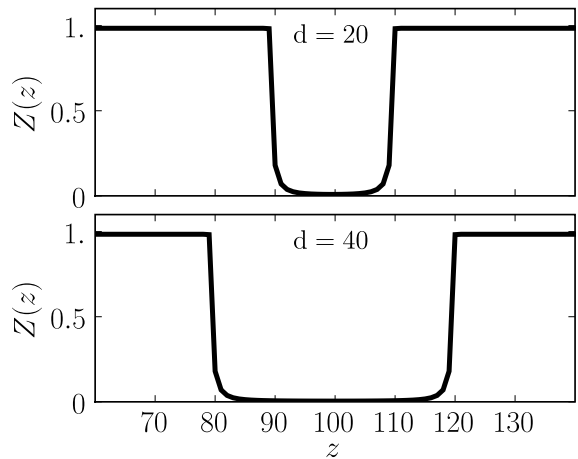}
\caption{Spatial dependence of $Z(z)$ for $U_{\rm left}=U_{\rm right} =2t$ and $U_{\rm center}= 15.9712t$.
The upper panel refers to a central region of $d=20$ layers, while the lower panel to $d=40$}\label{Fig_barrier_1}
\end{figure}
\begin{figure}[t]
\includegraphics[width=8cm]{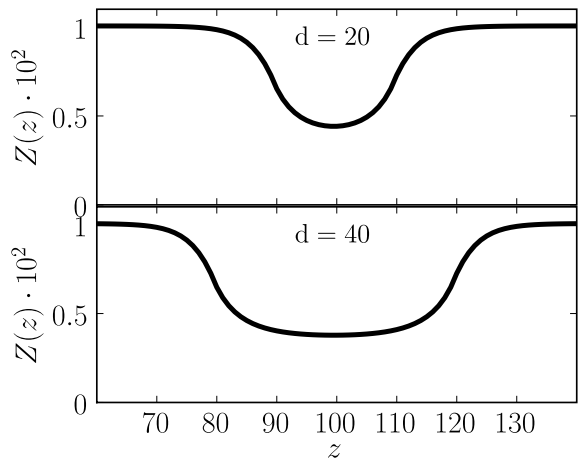}
\caption{Same as in Fig.~\ref{Fig_barrier_1}, for 
$U_{\rm left}=U_{\rm right} =15.9198t$ and $U_{\rm center}= 15.9712t$.\label{Fig_barrier_2}}
\end{figure}
\begin{figure}[t]
\includegraphics[width=8cm]{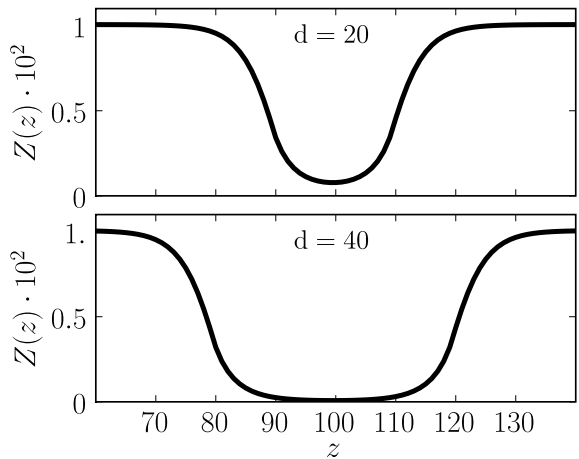}
\caption{Same as in Fig.~\ref{Fig_barrier_1}, for $U_{\rm left}=U_{\rm right} = 15.9198t$ and 
$U_{\rm center}= 16.0288t$.\label{Fig_barrier_3}}
\end{figure}
\begin{figure}[t]
\includegraphics[width=8cm]{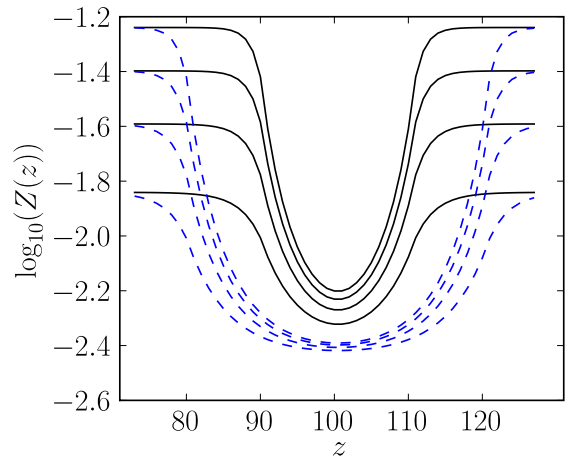}
\caption{(Color online) Logarithm of the quasiparticle weight $Z$ as a function of layer 
index $z$ for a 20-sites wide (solid line) and 40-sites wide 
(dashed line) strongly correlated metallic slab $U=15.9712t<U_{\rm crit}$ sandwiched between two weakly correlated metal leads 
(with $U=15.88438t$, $15.79388t$, $15.67674t$, $15.53236t$.). The entire system is 200-sites wide; the interfaces between 
the leads and the slab are at $z=80$ and $z=120$ for the 40-sites wide slab and $z=90$ and $z=110$ for the 20-sites wide slab. 
The figure shows that for increasing slab width the quasiparticle weight goes to a value that is independent of lead correlation.
}\label{Fig_barrier_logresults1}
\end{figure}
\begin{figure}[t]
\includegraphics[width=8cm]{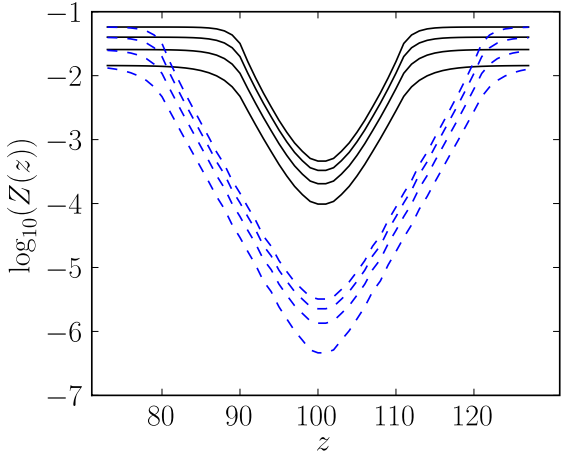}
\caption{(Color online) Same as in Fig.~\ref{Fig_barrier_logresults1}, but the central layers have now $U=16.1148>U_{\rm crit}$. 
In this case the quasiparticle weight at the center of the junction is strongly dependent both on barrier width and on the strength 
of electron correlation in the leads. The central layer remains metallic for arbitrary values of $U>U_{\rm crit}$, 
but its quasiparticle weight decreases exponentially with the slab width.}\label{Fig_barrier_logresults2}
\end{figure}

In this section we consider geometry (c) of figure \ref{Fig_geometries}, 
in which a strongly correlated slab of $d$ layers is sandwiched between two weakly correlated 
metal leads, a setup already studied by DMFT~\cite{Freericks,Pruschke-condmat}.
In Figs.~\ref{Fig_barrier_1}, \ref{Fig_barrier_2} and \ref{Fig_barrier_3} we show the layer dependence of the quasiparticle weight 
for different values of the interaction parameters, the Hubbard $U$ in the leads, $U_{\rm right}=U_{\rm left}<U_{\rm crit}$, 
and in the central slab, $U_{\rm center}\stackrel{\textstyle >}{<} U_{\rm crit}$, 
and slab thickness $d$. From those results one can draw the 
following conclusions:
\begin{itemize}
\item For any finite thickness $d$, the quasiparticle weight in the central slab never vanishes, as better revealed in 
Figs.~\ref{Fig_barrier_logresults1} and \ref{Fig_barrier_logresults2}, even for $U_{\rm center}>U_{\rm crit}$, 
fed as it is by the evanescent metallic quasiparticle strength from the metallic leads. This result agrees perfectly with 
recent DMFT calculations~\cite{Pruschke-condmat}.
\item For $U_{\rm center}>U_{\rm crit}$, see Fig.~\ref{Fig_barrier_3}, the minimum value $Z_{\rm min}$ in the central region 
decreases when $d$ increases;
\item The behavior of $Z(z)$ across the interface is smoother and smoother the closer and closer $U_{\rm right}=U_{\rm left}$ 
are to $U_{\rm center}$.
\end{itemize} 

Looking more in detail at Figs.~\ref{Fig_barrier_2},~\ref{Fig_barrier_3} and at the log-scale plots in 
Fig.~\ref{Fig_barrier_logresults1} and~\ref{Fig_barrier_logresults2}, we can identify the characteristic differences between a Mott insulating 
slab and a strongly correlated metallic slab, when sandwiched between metallic leads. In 
a strongly correlated metallic slab, the central quasiparticle weight ultimately settles to the self-standing value it would have in 
a homogeneous system with $U=U_{\rm center}<U_{\rm crit}$. 
This value is independent of the junction width and of lead correlations. 
On the contrary, the quasiparticle weight inside the insulating slab is completely borrowed from the leads, 
and strongly depends therefore on their separation and correlation. What depends strictly 
on the central slab interaction $U_{\rm center} > U_{\rm crit}$ is the quasiparticle decay 
length $\xi_{\rm center}$ from the lead to the center of the slab, which 
increases for increasing slab correlation according to the law $(U_{\rm center}-U_{\rm crit})^{-\nu}$, with $\nu\approx 0.5$, 
a value that matches perfectly that 
found in section~\ref{firstcase}

These considerations suggest that, if we look at the problem from a transport point of view, we are confronted with 
two completely different mechanisms. 
In a strongly correlated metallic central slab, $\xi_{\rm center}$ has the role of a screening length,  
exactly the same role of $\xi_{\rm right}$ in section~\ref{secondcase}. If instead the central slab is insulating, 
the meaning of $\xi_{\rm center}$ becomes completely different, it is now a tunneling length. No local quasiparticle 
peak would survive in a homogeneous Mott insulator: the residual quasiparticle peak that we find inside the central 
slab is therefore the evanescent lead electron wavefunction that tunnels into the slab.

A special case occurs when $U_{\rm center}\approx U_{\rm crit}$, i.e. right at criticality, where 
neither of the previous two pictures is valid. The crossover from the two opposite exponential decays describing either 
screening or tunneling is characterized by the absence of any characteristic length, which implies a power law variation 
of the quasiparticle strength upon the slab width $d$
\be
Z_{\rm min}(d) \sim \frac{1}{d^2} + O\left(\frac{1}{d^3}\right).\label{Z-d^2}
\ee
We find that the leading $1/d^2$ behavior is, within our accuracy, independent of the specific properties of the metallic leads, 
while the subleading terms do depend on them, 
see Fig.~\ref{Fig_scaling_mim_c}. 
A simple analytical justification of the critical $1/d^2$ behavior is provided in appendix~\ref{analytical}.

\begin{figure}[t]
\includegraphics[width=9.1cm]{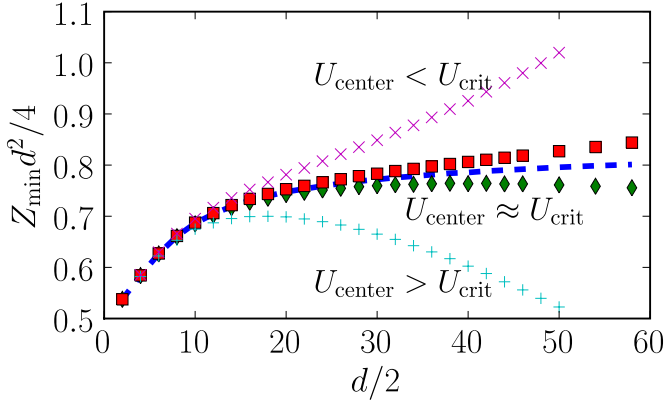}
\caption{(Color online) Numerical results for $Z_{\rm min}d^2/4$ and $U=15.999 t$ (crosses), 
$16 t$ (squares), $16.0002t$ (dashed line), 
$16.0004t$ (diamonds), $16.002t$ (pluses) for the sandwich geometry with $U_{\rm left} = U_{\rm right}=2t$. 
The constant value approached for $U=16.0002t\approx U_{\rm crit}$ and large junction width should be compared to the one we find 
in Eq.~\eqref{Z(l)-critical}.}\label{Fig_scaling_mim}
\end{figure}

\begin{figure}[t]
\includegraphics[width=9.1cm]{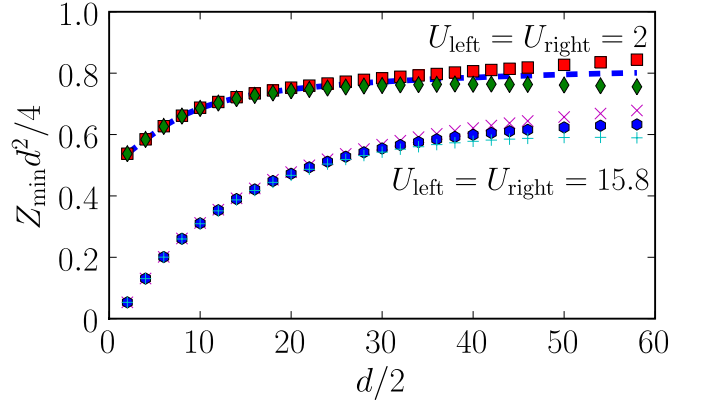}
\caption{(Color online) Numerical results for $Z_{\rm min}d^2/4$ for $U_{\rm left} = U_{\rm right}=2t$ 
[$U_{\rm center}=16 t$ (squares), $16.0002t$ (dashed line), $16.0004t$ (diamonds)], and for $U_{\rm left} = U_{\rm right}=15.8t$ 
[$U_{\rm center}=16.0002 t$ (crosses), $16.0004t$ (hexagons), $16.0006t$ (pluses)]. The stronger lead 
correlation in the lower curves pushes the plateau of the function $Z_{\rm min}d^2/4$ towards larger values of $d$.}
\label{Fig_scaling_mim_c}
\end{figure}

\section{Conclusions}\label{Conclusions}

In this work we have studied how the spatial inhomogeneity of interfaces affects the physics of a strongly correlated electron 
system. To address this problem, we extended the conventional Gutzwiller approximation technique to account for 
inhomogeneous Hamiltonian parameters. Moreover, to efficiently cope with the larger number of variational parameters in comparison 
with the homogeneous case, we derived iterative equations fully equivalent to the saddle point equations that identify 
the optimal variational solution, similarly to what is 
commonly done within unrestricted Hartree-Fock or {\sl ab initio} LDA calculations. These iterative equations can be solved 
without much effort for very large system sizes; an advantage with respect to more rigorous approaches, like e.g. DMFT  
calculations, which are numerically feasible only for small systems. 

We have applied the method to various interface geometries in three dimensions; specifically the interface of a strongly correlated 
metal with the vacuum, the interface between two differently correlated metals and the junction between two weakly correlated 
metals sandwiched by a strongly correlated slab. All these geometries had been already studied by 
DMFT~\cite{Freericks,Freericks-PRB-07,Potthoff-2,Potthoff-1,Liebsch,ishida,Rosch,Pruschke-condmat}, which 
allowed us to directly compare our results with more rigorous ones, thus providing a test on the quality of our 
approximation, which is then applied to much larger sizes.

Our main result is that the effects of an interface decay exponentially in the interior of a strongly 
correlated system on a very long length-scale proportional to the correlation length of the incipient Mott transition, 
a bulk property independent upon the details of the interface.~\cite{Ours} In particular, at the surface of a strongly 
correlated metal we find a strong suppression of the metallic properties, e.g. of the quasiparticle weight, 
that persists on a large depth controlled by the Mott transition correlation length, a ``dead layer''~\cite{Ours} appearing because 
the surface is effectively more correlated than the bulk and consistent with photoemission experiments.~\cite{Marsi} 
Conversely, metallic features from a metal lead penetrate inside a Mott insulator within a depth that, once again, diverges 
on approaching the Mott transition. As a consequence, a conducting channel always exists inside a Mott insulating slab 
contacted to two metallic leads, in agreement with recent DMFT analyses~\cite{Pruschke-condmat}, implying a finite conductance 
at zero bias and temperature 
that decays fast on increasing both external parameters on an energy scale exponentially small in the length of the slab in units of  
the Mott transition correlation length.  

The method that we have developed is very simple and flexible, so it can in principle be applied to a variety of  
realistic situations of current interest, not only for studying interfaces but also for more general inhomogeneities, as those 
arising by impurities or other defects, and can easily incorporate additional features like magnetism, which we have disregarded 
throughout this work.   

\begin{acknowledgments}
The work was supported by the Italian Ministry of University
and Research, through a PRIN-COFIN award. The environment provided by the
independent ESF project CNR-FANAS-AFRI was also useful.
\end{acknowledgments}

\appendix
\section{Analytical expressions near criticality}\label{analytical}

\begin{figure}[bht]
\includegraphics[width=8.1cm]{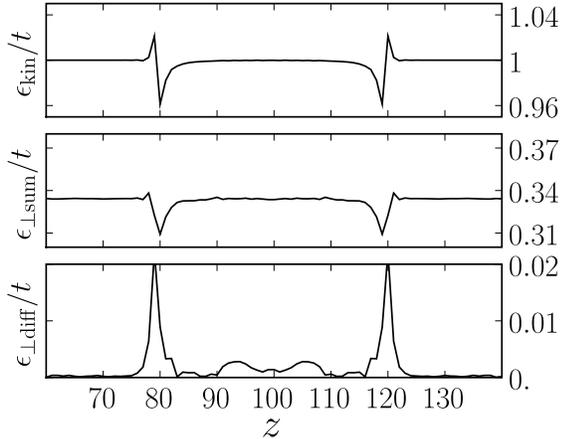}
\caption{Upper panel, plot $\epsilon_{\rm kin}/t$ for the sandwich geometry (c) 
with 40 central layers, $U_{\rm left}=U_{\rm right} = 2t$ and $U_{\rm center} = 15.9712t$. 
The value deviates by 2 to 4\% from the value it would have in a homogeneous system 
($\tilde{\varepsilon}_{\rm kin}=t$). Middle panel, plot of 
$\epsilon_{\perp {\rm sum}}=\epsilon_{\perp}(z+1/2)+\epsilon_{\perp}(z-1/2)$. 
Lower panel, plot of $\epsilon_{\perp {\rm diff}}=\epsilon_{\perp}(z+1/2)-\epsilon_{\perp}(z-1/2)$}\label{Fig_scaling_mim_sum}
\end{figure}

In this appendix, we show how to derive simple analytical expressions for the layer dependence of the quasiparticle residue 
near criticality. We assume a three dimensional slab geometry with constant hopping but inhomogeneous interaction $U(z)$ and  
with particle-hole symmetry. We define as $2\epsilon_{||}(z)$ and $2\epsilon_\perp(z-1/2)$ the average over the uncorrelated Slater 
determinant $|\Psi_0\rangle$ of the hopping energy per bond within layer $z$ and between layers $z$ and $z-1$, respectively. 
With these definitions, the equation~\eqn{due} can be written as
\begin{widetext}
\bea
0 &=&  2\,R(z)\bigg(4\,\epsilon_{||}(z) + \epsilon_\perp(z-1/2) + \epsilon_\perp(z+1/2)\bigg) 
+ \bigg(\epsilon_\perp(z-1/2)+\epsilon_\perp(z+1/2)\bigg)\bigg(R(z+1)+R(z-1)-2R(z)\bigg)\nonumber\\
&& + \bigg(\epsilon_\perp(z+1/2)-\epsilon_\perp(z-1/2)\bigg)\bigg(R(z+1)-R(z-1)\bigg)
+ \fract{U(z)}{4}\,\fract{R(z)}{\sqrt{1-R^2(z)}}.\label{due-bis}
\eea
\end{widetext}
Near criticality, we expect that the layer dependence must appear as a dependence upon the scaling variable $z/\xi$, and,  
since $\xi\gg 1$, it becomes allowed to regard $z/\xi$ as a continuous variable and expand \eqn{due-bis} in the leading gradients. 
Because of the interface, both $\epsilon_{||}(z)$ and $\epsilon_\perp(z-1/2)$ must acquire a Friedel-like $z$-dependence. 
However, as shown explicitly in Fig.~\ref{Fig_scaling_mim_sum}, $\epsilon_{||}(z)$ and  
$\epsilon_\perp(z-1/2) + \epsilon_\perp(z-1/2)$ vary appreciably only close to the interfaces, while 
$\epsilon_\perp(z-1/2) - \epsilon_\perp(z-1/2)$ is negligible. Indeed, as 
discussed in more detail in the Appendix~\ref{friedel}, 
the amplitude of the Friedel's oscillations is strongly reduced near criticality, while the period stays invariant, so that 
it is legitimate to neglect the $z$ dependence of 
$\epsilon_{||}(z)$ and $\epsilon_\perp(z\pm 1/2)$ and use for them their large-$z$ bulk values, 
$\epsilon_{||}$ and $\epsilon_\perp$. 

Noting that the average hopping energy per site in the homogeneous case 
is $\epsilon_{kin} = 4\epsilon_{||}+2\epsilon_{\perp}$, the above Eq.~\eqn{due-bis} can be written in the continuous limit as    
\be
2\,R(z)\,\epsilon_{kin} + \fract{U}{4}\,\fract{R(z)}{\sqrt{1-R^2(z)}} + 
2\epsilon_\perp\,\fract{\partial^2 R(z)}{\partial z^2} = 0,\label{due-ter}
\ee
where we take the bulk value $U(z)=U$, since its variation is limited to a single layer. Eq.~\eqn{due-ter} admits 
an integral of motion, namely
\bea
E &=&\epsilon_{\perp}\,\left(\fract{\partial R(z)}{\partial z}\right)^2 + 
\epsilon_{kin}\,R^2(z) \nonumber \\
&&~~~~ + \fract{U}{4}\left(1-\sqrt{1-R^2(z)}\right) \nonumber \\
&& \equiv \epsilon_{\perp}\,\left(\fract{\partial R(z)}{\partial z}\right)^2 + E\left[R(z)\right], \label{due-quater}
\eea
where $E\left[R(z)\right]$ is the Gutzwiller variational energy for a homogeneous system 
calculated at fixed $R=R(z)$, i.e. not the optimized one. The constant of motion $E$ must be chosen to correspond to 
$E[R(z_0)]=E[R_0]$, where $z_0$ is the layer coordinate at which we expect vanishing derivative. In a single interface, 
we expect that $R(z)$ will reach a constant value only asymptotically far from the interface, i.e. $z_0\to\infty$, where  
$R_0$ tends to its bulk value
\[
R_0 = \sqrt{1-u^2},
\] 
and $E[R_0]$ to the optimized energy in a homogeneous system, i.e.  
\[
E = E[R_0] = -\fract{U_{\rm crit}}{8}\left(1-u\right)^2\,\theta(1-u),
\]
with $u=U/U_{\rm crit}$ and $U_{\rm crit} = -8\epsilon_{kin}$, in the Gutzwiller approximation. 
In the case of a correlated slab sandwiched between two metal leads, we expect that $R(z)$ will reach a minimum 
somewhere at midway between the two interfaces. If the leads are identical, the minimum occurs right in the middle, so that 
$R_0$ becomes an unknown parameter that has to be fixed by imposing that the actual solution $R[z,R_0]$, which depends parametrically 
on $R_0$, has a vanishing slope $\partial_z\,R[z,R_0] = 0$ for $z$ in the middle of the slab.     

With the same definitions as above, 
\[
E\left[R(z)\right] = - \fract{U_{\rm crit}}{8}\,R^2(z) + \fract{U_{\rm crit}}{4}\, u\, \left(1-\sqrt{1-R^2(z)}\right).
\]
Since in a homogeneous cubic lattice $\epsilon_\perp = \epsilon_{kin}/6 = - U_{\rm crit}/48$, Eq.~\eqn{due-quater} 
can be rewritten as 
\bea
&& \frac{1}{6}\,\left(\fract{\partial R(z)}{\partial z}\right)^2 = 
R_0^2 +2u\,\left(1-\sqrt{1-R_0^2}\right)\nonumber \\
&&~~~~~~~ - R^2(z) +2u\,\left(1-\sqrt{1-R^2(z)}\right),\label{Eq}
\eea  
where 
\be
R_0^2 +2u\,\left(1-\sqrt{1-R_0^2}\right) = \left(1-u\right)^2\, \theta(1-u),\label{E-single-interface}
\ee
in the case of a single interface. The pre-factor 6 of the $(\partial R(z)/\partial z)^2$ comes from the 
homogeneous relation $\epsilon_{kin}/\epsilon_\perp =6$. As we shall see, the numerical data can be better interpreted if 
$\epsilon_{kin}/\epsilon_\perp$ is considered as a free fitting parameter 

The differential equation \eqn{Eq} controls the $z$-dependence of $R(z>0)$, hence of the quasiparticle residue $Z(z)=R^2(z)$, 
assuming that the interface affects only the boundary condition $R(z=0) = R_{\rm surf}$. Therefore, a surface less correlated 
than the bulk should be described by \eqn{Eq} with $R_{\rm surf} > R_{\rm bulk}=\sqrt{1-u^2}\,\theta(1-u)$, 
while by $R_{\rm surf} < R_{\rm bulk}$ the opposite case, as for instance the interface with the vacuum of section~\ref{firstcase}. 

We now consider separately the case of a single junction and of the double junction, with either metallic or insulating bulk. 

\subsection{Single interface with metallic bulk: $u\leq 1$}\label{analytical_int_metallic}

In the case of a single interface, Eq.~\eqn{E-single-interface} with $u\leq 1$ has to be used.  
The differential equation \eqn{Eq} reads
\[
\frac{1}{6}\,\left(\fract{\partial R(z)}{\partial z}\right)^2 = \bigg(\sqrt{1-R^2(z)} - u\bigg)^2,
\]
hence 
\[
\fract{\partial R(z)}{\partial z} = \sqrt{6}\,\bigg(\sqrt{1-R^2(z)} - u\bigg), 
\]
namely 
\[
\int_{R_{\rm surf}}^{R(z)} \fract{dR}{\sqrt{1-R^2} - u} = \sqrt{6}\,z. 
\]
This integral equation can be solved exactly, leading to the implicit formula
\ba
\sqrt{6}\,z &=& \int_{\arcsin R_{\rm surf}}^{\arcsin R(z)} \fract{\cos x\, dx}{\cos x - u} \\
&=&  \arcsin R(z) - \arcsin R_{\rm surf} \\
&&\!\!\!\!\!\!\!\!\!\! + \fract{u}{\sqrt{1-u^2}}\,\tanh^{-1}\left(\fract{R(z)\,R_{\rm bulk}}{1-\sqrt{\left(1-R_{\rm bulk}^2\right)
\left(1-R^2(z)\right)}}
\right)\\
&&\!\!\!\!\!\!\!\!\!\! 
- \fract{u}{\sqrt{1-u^2}}\,\tanh^{-1}\left(\fract{R_{\rm surf}\,R_{\rm bulk}}{1-\sqrt{\left(1-R_{\rm bulk}^2\right)
\left(1-R_{\rm surf}^2\right)}}\right).  
\ea
Close to criticality, $u\simeq 1$, one can neglect the arcsines in the rhs and find the explicit expression 
\be
R(z) = \fract{R_{\rm bulk}\,\sinh \zeta}{\cosh\zeta \pm \sqrt{1-R^2_{\rm bulk}}},\label{R(z)-u<1}
\ee
where the plus sign refers to the case $R_{\rm surf}<R_{\rm bulk}$, and the minus sign to the opposite case, and 
\bea 
\zeta &=& \sqrt{6\left(1-u^2\right)}\;z  \nonumber \\
&& \!\!\!\!\!\!\!\!+ 
\tanh^{-1}\left(\fract{R_{\rm surf}\,R_{\rm bulk}}{1-\sqrt{\left(1-R_{\rm bulk}^2\right)
\left(1-R_{\rm surf}^2\right)}}\right)\nonumber\\
&& \equiv \sqrt{6}\;R_{\rm bulk}\,\left(z + z_*\right).\label{zeta-u<1}
\eea
This solution provides a definition of the correlation length for $u\alt 1$
\be
\xi = \fract{1}{\sqrt{6\left(1-u^2\right)}}\simeq 0.289\,\left(\fract{U_{\rm crit}}{U_{\rm crit} - U}\right)^{1/2},
\label{xi-u<1}
\ee
quite close to the DMFT value.~\cite{Rosch} We note that, for $\zeta\gg 1$, Eq.~\eqn{R(z)-u<1} becomes
\[
R(z\to\infty)\simeq R_{\rm bulk}\,\left(1\mp \sqrt{1-R_{\rm bulk}^2}\; \mathrm{e}^{-\zeta}\right),
\]
therefore
\be
Z(z) = R^2(z)\simeq Z_{\rm bulk}\,\left(1 \mp 2\sqrt{1-R_{\rm bulk}^2}\;\mathrm{e}^{-x/\xi}\right),\label{approach-bulk}
\ee
tends exponentially to its bulk value on a length scale $\xi$, from below or above according to 
$R_{\rm surf}\stackrel{\textstyle <}{>} R_{\rm bulk}$, respectively. 

\begin{figure}[htb]
\includegraphics[width=8.cm]{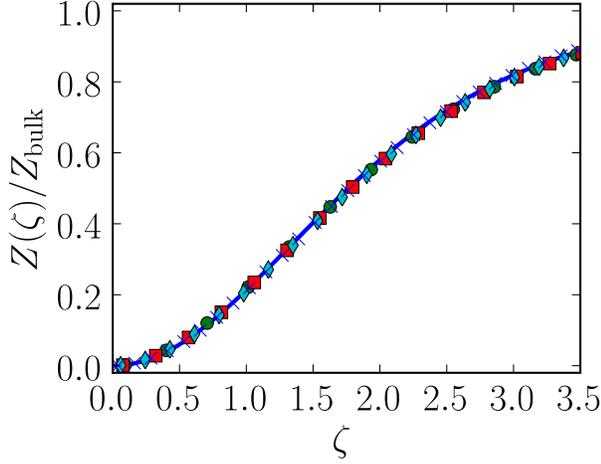}
\caption{(Color online) Numerical results for $Z(z)$ in the surface geometry, 
with $U=15.9872t$ (crosses), $15.9712t$ (diamonds), $15.9487t$ 
(squares), $15.9198t$ (circles). The solid curve is $\tanh^2(\zeta/2)$, i.e. $R^2(\zeta)$ as defined in Eq.~\eqref{R(z)-u<1} 
(with plus sign) and expanded to first order in $R_{\rm bulk} \ll 1$. In order to define $\zeta$ the same expansion has been carried 
out in Eq.~\eqref{zeta-u<1}, where we set the quantity $\epsilon_{kin}/\epsilon_{\perp}$ equal to 9.427 instead of 6, in order 
to fit the numerical data.}\label{Fig_scaling_surf}
\end{figure}

Near criticality, i.e. $R_{\rm bulk}=\sqrt{1-u^2}\ll 1$, Eq.~\eqn{R(z)-u<1} becomes 
\be
R(z)\simeq R_{\rm bulk}\,\coth\frac{\zeta}{2},\label{R(z)-u<1:a}
\ee
so that
\begin{align}
\left(z+z_*\right)^2\,Z(z) &= \left(z+z_*\right)^2\,R(z)^2 \nonumber \\
&= \frac{4}{6}\,\bigg(\frac{1}{4}\,\zeta^2\,\coth^2\frac{\zeta}{2}\bigg)
\equiv \frac{2}{3}\,f_{u<1}(\zeta),\label{intercept-u<1}
\end{align}
shows a simple scaling behavior~\cite{Rosch}. The scaling function $f_{u<1}(\zeta)$ that we find has the 
asymptotic behavior: $f_{u<1}(0)=1$ and $f_{u<1}(\zeta\to\infty)\simeq \zeta^2/4$. 

Another case of interest is that of the interface with vacuum discussed in section~\ref{firstcase}. Here $R_{\rm surf}\ll 1$
hence from Eq.~\eqn{zeta-u<1} it follows that 
\[
z_* \simeq \fract{R_{\rm surf}}{\sqrt{6} (1-u)}\ll 1.
\]
Away from criticality and for $\zeta\ll 1$, which is allowed since 
$z_*\ll 1$, we find through \eqn{R(z)-u<1} with the plus sign that 
\[
R(z) \simeq \sqrt{6}\,\left(1-u\right)\,\left(z+z_*\right),
\]
so that 
\be
Z(z) \simeq 6\,\left(1-u\right)^2\,\left(z+z_*\right)^2,\label{Z(z)-small-z}
\ee
showing that the quasiparticle residue approaches its surface value with a finite curvature.

\begin{figure}[hbt]
\includegraphics[width=7.5cm]{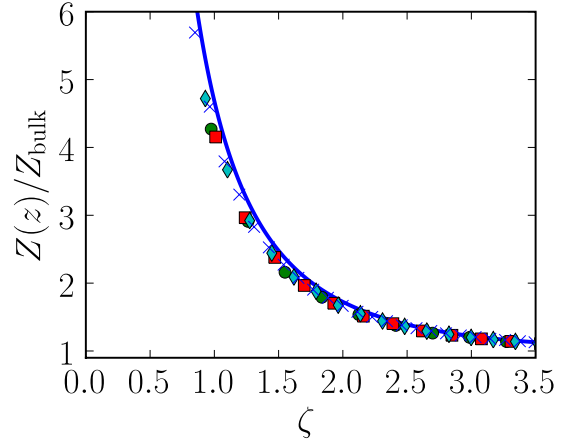}
\caption{(Color online) Numerical results for $Z(z)$ in the single junction geometry with metallic bulk, 
the position of the junction is 
chosen as the origin for the spatial coordinate, the metal on the left side is very weakly correlated ($U=2t$); the values 
for $U$ on the right side are the same of Fig.~\ref{Fig_scaling_surf}. The solid curve is now the function $1/\tanh^2(\zeta/2)$, 
i.e. the second power of Eq.~\eqref{R(z)-u<1} (with minus sign) expanded to first order in $R_{\rm bulk}$. 
As in Fig.~\ref{Fig_scaling_surf}, the definition of $\zeta$ has been obtained from Eq.~\eqref{zeta-u<1} by expanding to 
first order in $R_{\rm bulk}$. The value of $\epsilon_{kin}/\epsilon_{\perp}$ that fits the data is now 8.254.}\label{Fig_scaling_wcmscm}
\end{figure}

In Fig.~\ref{Fig_scaling_surf} and Fig.~\ref{Fig_scaling_wcmscm} we show that rescaled numerical data for an 
interface between a 200-layer-wide correlated metal slab and the vacuum and for a junction between a 
weakly correlated metal and a strongly correlated metal. It is easy to fit the numerical data with the function 
$R^2(z)$ displayed in Eq.~\eqref{zeta-u<1} by tuning just one parameter, which, as discussed above, is the value 
of $\epsilon_{kin}/\epsilon_{\perp}$ 
(equal to 6 in the homogeneous problem). The fact that the ideal theoretical result, relying on homogeneous values for hopping and 
kinetic energy, fits the numerical data with just a single tunable parameter, is a pleasant feature.

\subsection{Single interface with insulating bulk: $u\geq 1$} 

In this case the equation \eqn{Eq} using \eqn{E-single-interface} with $u\geq 1$ reads
\be
\frac{1}{6}\,\left(\fract{\partial R(z)}{\partial z}\right)^2 = - R^2(z) +2u\,\left(1-\sqrt{1-R^2(z)}\right),\label{Eq-1}
\ee  
leading to 
\[
\int \fract{dR}{\sqrt{2u -R^2 - 2u\sqrt{1-R^2}}} = -\sqrt{6}\int dz,
\]
where we have assumed that on the surface $R_{\rm surf}$ is finite and decay in the bulk, so that the derivative is negative.  
The above integral equation can be solved too, with an implicit solution 
\begin{widetext}
\ba
-\sqrt{6\left(u-1\right)}\, z &=& 2\,\sqrt{u-1}\,\arcsin\left(\fract{\cos y(z)}{\sqrt{u}}\right)
- 2\,\sqrt{u-1}\,\arcsin\left(\fract{\cos y_{\rm surf}}{\sqrt{u}}\right)\\
&& - \tanh^{-1}\left(\fract{\sqrt{u-1}\,\cos y(z)}{\sqrt{u-\cos^2 y(z)}}\right)
+ \tanh^{-1}\left(\fract{\sqrt{u-1}\,\cos y_{\rm surf}}{\sqrt{u-\cos^2 y_{\rm surf}}}\right)
\ea
\end{widetext}
where $R(z) =\sin 2y(z)$, $R_{\rm surf}=\sin 2y_{\rm surf}$. As before the arcsines can be neglected near criticality to 
obtain the explicit solution  
\be
R^2(z) = 1 - \left(1-\fract{2\left(u-1\right)}{u\cosh^2\zeta - 1}\right)^2,\label{R(z)-u>1}
\ee
with 
\bea
\zeta &=& \sqrt{6\left(u-1\right)}\, z + \tanh^{-1}\left(\fract{\sqrt{u-1}\,\cos y_{\rm surf}}{\sqrt{u-\cos^2 y_{\rm surf}}}\right)
\nonumber\\
&&\equiv \sqrt{6\left(u-1\right)}\,\left(z+z_*\right). 
\label{zeta-u>1}
\eea
In the case of an insulating bulk, the correlation length defined through \eqn{xi-u>1} is therefore 
\be
\xi = \fract{1}{\sqrt{6\left(u-1\right)}} \simeq 0.408 \, 
\left(\fract{U_{\rm crit}}{U-U_{\rm crit}}\right)^{1/2},\label{xi-u>1}
\ee
with a different numerical prefactor, actually a $\sqrt{2}$ greater, with respect to the metallic bulk \eqn{xi-u<1}. 

Near criticality, $u\agt 1$, 
\[
R(z)^2 =Z(z) \simeq \fract{4\left(u-1\right)}{\sinh^2 \zeta},
\]  
so that, as before, 
\bea
\left(z+z_*\right)^2\,Z(z) &=& \fract{4}{6}\left(\fract{\zeta^2}{\sinh^2 \zeta}\right)\nonumber\\
&&\equiv \frac{2}{3}\,f_{u>1}(\zeta),\label{intercept-u>1}
\eea
has a scaling behavior with $f_{u>1}(0)=1$ and 
\[
f_{u>1}(\zeta\to\infty)\simeq 4\zeta^2\,\mathrm{e}^{-2\zeta}.
\]

\subsection{Double junction}

We assume for simplicity a slab of length $2L$ in contact with two leads. 
In this case we need to use Eq.~\eqn{Eq} with $R_0$ a parameter that has to be fixed by imposing that the solution 
$R(z)$ becomes $R_0$ at some $z_0$ within the slab. If we assume that both leads are less correlated than the slab, then $R(z)$ 
always decreases moving away from any of the two interfaces, and we can determine $R_0$ by imposing either of the two 
following conditions:
\begin{widetext}
\bea 
\int_{R^<_{\rm surf}}^{R_0} \fract{dR}{\sqrt{R_0^2+2u\sqrt{1-R_0^2}-R^2-2u\sqrt{1-R^2}}}
&=& - \sqrt{6}\,z_0,\label{left-surface}\\
\int_{R_0}^{R^>_{\rm surf}} \fract{dR}{\sqrt{R_0^2+2u\sqrt{1-R_0^2}-R^2-2u\sqrt{1-R^2}}}
&=& \sqrt{6}\,\left(2L-z_0\right),\label{right-surface},
\eea
\end{widetext}
where $R^<_{\rm surf}$ and $R^>_{\rm surf}$ are the values of $R(z)$ at the left and right surfaces, respectively. 
Taking the difference \eqn{right-surface} minus \eqn{left-surface} we find 
\begin{widetext}
\be
\sqrt{6}\,2L =  \left(\int_{R_0}^{R^>_{\rm surf}} + \int_{R_0}^{R^<_{\rm surf}}\right)\,  
\fract{dR}{\sqrt{R_0^2+2u\sqrt{1-R_0^2}-R^2-2u\sqrt{1-R^2}}},\label{Eq-sandwich}
\ee
\end{widetext}
which has to be solved to find $R_0$ as function of the other parameters. Once $R_0$ is found, one can determine $z_0$. 
In order to simplify the calculations, we will assume two identical leads, i.e. 
$R^<_{\rm surf} = R^>_{\rm surf} = R_{\rm surf}$, so that $z_0=L$ and \eqn{Eq-sandwich} becomes 
\begin{widetext}
\be
\sqrt{6}\, L = \int_{R_0}^{R_{\rm surf}} \fract{dR}{\sqrt{R_0^2+2u\sqrt{1-R_0^2}-R^2-2u\sqrt{1-R^2}}}
= \fract{2}{\sqrt{(a-c)(b-d)}}\left[
(c-b)\,\Pi\left(\phi,\frac{c-d}{b-d},k\right) + b\,F(\phi,k)\right],\label{sandwich-solution}
\ee
\end{widetext}
with parameters $a>b>c>u\geq d$. The last expression can be derived easily after the change of variable $R=\sqrt{1-x^2}$, and 
seemingly $R_0=\sqrt{1-x_0^2}$ and $R_{\rm surf} = \sqrt{1-x_{\rm surf}^2}$. 
$\Pi(\phi,n,k)$ and $F(\phi,k)$ are elliptic integrals of third and first kind, respectively
\ba
F(\phi,k) &=& \int_0^\phi \fract{dx}{\sqrt{1-k^2\sin^2 x}},\\
\Pi(\phi,n,k) &=& \int_0^\phi \fract{dx}{\left(1-n\sin^2 x\right)\sqrt{1-k^2\sin^2 x}},
\ea 
and 
\ba
\phi &=& \arcsin\sqrt{\fract{(b-d)(c-u)}{(c-d)(b-u)}},\\
k &=& \sqrt{\fract{(a-b)(c-d)}{(a-c)(b-d)}}.
\ea
The various parameters are, when $2u-x_0\geq 1$, 
\ba
a &=& 2u-x_0,\\
b &=& 1,\\
c &=& x_0,\\
d &=& -1,\\
u &=& x_{\rm surf},
\ea
so that 
\ba
\phi &=& \arcsin\sqrt{\fract{2\left(x_0-x_{\rm surf}\right)}{\left(x_0+1\right)\left(1-x_{\rm surf}\right)}},\\
k &=& \sqrt{\fract{\left(2u-x_0-1\right)\left(x_0+1\right)}{4\left(u-x_0\right)}}.
\ea
On the contrary, if $2u-x_0 < 1$, then 
\ba
a &=& 1,\\
b &=& 2u-x_0,\\
c &=& x_0,\\
d &=& -1,\\
u &=& x_{\rm surf},
\ea
hence
\ba
\phi &=& \arcsin \sqrt{\fract{(2u-x_0+1)(x_0-x_{\rm surf})}{(x_0+1)(2u-x_0-x_{\rm surf})}},\\
k &=& \sqrt{\fract{(1-2u+x_0)(x_0+1)}{(1-x_0)(2u-x_0+1)}}.
\ea
We rewrite 
\bea
&& (c-b)\,\Pi\left(\phi,\frac{c-d}{b-d},k\right) + b\,F(\phi,k)\label{express-1} \\
&& = \int_0^\phi dx\; \left(\fract{d(b-c)+b(c-d)\cos^2 x}{(b-c)+(c-d)\cos^2 x}\right)\;
\fract{1}{\sqrt{1-k^2\sin^2 x}},\nonumber
\eea
and note that at $x=\phi$
\[
\fract{d(b-c)+b(c-d)\cos^2 \phi}{(b-c)+(c-d)\cos^2 \phi} = x_{\rm surf}\geq 0.
\]
In addition $b-c$ in both cases is very small. Indeed, for $2u-x_0>1$, which corresponds to an insulating slab where 
$R_0=\sqrt{1-x_0^2}\to 0$ for large $L$,  
$b-c=1-x_0\ll 1$. In the opposite case of a weakly correlated slab, still 
$b-c=2u-x_0-x_0\ll 1$ since $x_0\to u$ for large $L$. 
Therefore 
\[
\fract{d(b-c)+b(c-d)\cos^2 x}{(b-c)+(c-d)\cos^2 x} 
\]
is practically constant and equal to $b$ everywhere but close to the extreme of integration, where it fastly decays to $x_{\rm surf}$. 
Therefore to leading order we can write 
\[
(c-b)\,\Pi\left(\phi,\frac{c-d}{b-d},k\right) + b\,F(\phi,k) \simeq b\,F(\phi,k),
\]
hence the equation to be solved becomes 
\begin{widetext}
\be
\sqrt{6}\,L = \fract{2b}{\sqrt{(a-c)(b-d)}}\; F(\phi,k) 
= \fract{2b}{\sqrt{(a-c)(b-d)}}\;\left[K(k) - 
F\left(\arcsin \fract{\cos \phi}{\sqrt{1-k^2\sin^2 \phi}},k\right)\right],\label{eq-to-solve}
\ee
\end{widetext}
where $K(k)=F(\pi/2,k)$ and the last expression being more convenient since $\phi\simeq \pi/2$. 

In order to find $x_0$ as function of the other parameters, we have to consider separately three different cases.

\subsubsection{Insulating off-critical behavior: $u\gg 1$}
In this case $2u-x_0>1$. We note that $k$ as a function of $u$ at fixed $x_0\simeq 1$ is equal to 
\[
k^2 = \fract{x_0+1}{4}\simeq \frac{1}{2},
\]
for $u=1$, and very rapidly increases to its asymptotic $u\gg 1$ value
\[
k^2 = \fract{x_0+1}{2}\simeq 1.
\]
Therefore \eqn{eq-to-solve} is, at leading order,   
\[
\sqrt{6}\; L = \fract{1}{\sqrt{u-1}}\; K\left(\sqrt{\fract{1+x_0}{2}}\right)
\simeq \fract{1}{2\sqrt{u-1}}\;\ln \fract{32}{1-x_0}.
\]
Therefore, in this limit, 
\be
Z_0 = R^2_0 \simeq 64\;\mathrm{e}^{-\sqrt{24(u-1)}\;L},\label{Z(l)-off-critical}
\ee
vanishes exponentially in the length of the slab. 

\subsubsection{Critical behavior: $u=1$}
In this case 
\[
k^2 = \fract{x_0+1}{4}\simeq \frac{1}{2},
\]
hence at leading order Eq.~\eqn{eq-to-solve} reads  
\[
\sqrt{6}\;L = \fract{1}{\sqrt{1-x_0}}\,K\left(\fract{1}{\sqrt{2}}\right) = \fract{1}{4\sqrt{\pi}\,\sqrt{1-x_0}}\,
\left[\Gamma\left(\frac{1}{4}\right)\right]^2,
\]
from which it follows that  
\be
Z_0 = R_0^2 = \frac{1}{48 \pi}\,\left[\Gamma\left(\frac{1}{4}\right)\right]^4\;\frac{1}{L^2}
\simeq \fract{1.146}{L^2}.\label{Z(l)-critical}
\ee
Once again we find a critical behavior $L^2 Z_0 \simeq const.$, with a sizable constant 1.146.

\subsubsection{Metallic off-critical behavior: $u\ll 1$}

This is the case in which $2u-x_0<1$ and $x_0 \simeq u$, so that 
\[
1-k^2 = \fract{4(u-x_0)}{(2u-x_1+1)(1-x_0)}\simeq \fract{4(u-x_0)}{1-u^2}.
\]
Therefore Eq.~\eqn{eq-to-solve} is
\[
\sqrt{6}\;L \simeq \fract{u}{\sqrt{1-u^2}}\,\ln \fract{16}{1-k^2} = 
\fract{u}{\sqrt{1-u^2}}\,\ln \fract{4(1-u^2)}{u-x_0},
\]
whose solution is
\[
u-x_0 = 4(1-u^2)\;\mathrm{e}^{-\sqrt{6}\,\sqrt{1-u^2}\,L/u}.
\]
Therefore, since $Z_{\rm bulk} = 1-u^2$, it follows that 
\be
Z_0 \simeq Z_{\rm bulk}\left( 1 + 8u\;\mathrm{e}^{-\sqrt{6}\,\sqrt{1-u^2}\,L/u}\right).\label{Z(l)-off-critical-metal}
\ee

\subsection{Comparison with DMFT}

Near the Mott transition, $U\simeq U_{\rm crit}$, Potthoff and Nolting in Ref.~\onlinecite{Potthoff-2} have introduced a set 
of linearized DMFT recursive equations for the layer dependent quasiparticle residue. Taking, as before, the 
continuous limit of their Eq.~(37), with $q=4$ $p=1$ and $U_{\rm crit}=6t\sqrt{6}$, one finds the following differential 
equation
\be
\frac{1}{6}\,\fract{\partial^2 Z(z)}{\partial z^2} 
+ 2\,Z(z)\,\left(1-u\right) - c\,Z(z)^2 = 0.\label{Potthoff-eq}
\ee
The numerical constant is estimated to be $c=11/9$~\cite{Bulla&Potthoff}. 
The limiting behavior for $u\to 1$ is the solution of 
\[
\frac{1}{6}\,\fract{\partial^2 Z(z)}{\partial z^2} = c\,Z(z)^2,
\]
namely
\be
z^2\,Z(z) = \frac{1}{c}=\frac{9}{11}\simeq 0.82.\label{intercept-Potthoff}
\ee

Let's consider instead our Eq.~\eqn{due-ter} that, divided by $2\epsilon_{kin}=-U_{\rm crit}/4$, can be written as 
\bea
0 &=& \frac{1}{6}\,\fract{\partial^2 R(z)}{\partial z^2} + R(z) - u\,\fract{R(z)}{\sqrt{1-R(z)^2}}\nonumber\\ 
&& \simeq \frac{1}{6}\,\fract{\partial^2 R(z)}{\partial z^2} + \left(1-u\right)\,R(z) - \frac{1}{2}\,R(z)^3.\label{our-eq}
\eea
At criticality, $u\to 1$, the solution 
\be
z^2\, R(z)^2 = z^2\, Z(z) = \frac{2}{3}\simeq 0.66,\label{intercept-our}
\ee
is just the limiting value of Eqs.~\eqn{intercept-u>1} and \eqn{intercept-u<1} for $\zeta=0$. The numerical coefficient $2/3$ 
that we find is slightly smaller than the linearized DMFT one, $9/11$, but both are much bigger than the value 
extracted by straight DMFT calculations in Ref.~\onlinecite{Rosch}, namely $0.008$. Supposedly, straight DMFT is 
a better approximation than linearized DMFT, which in turns should be better than our Gutzwiller technique, therefore 
it is likely that our results overestimate the quasiparticle residue $Z$.

\section{Friedel's Oscillations}\label{friedel}

\begin{figure}[t!]
\includegraphics[width=8.7cm]{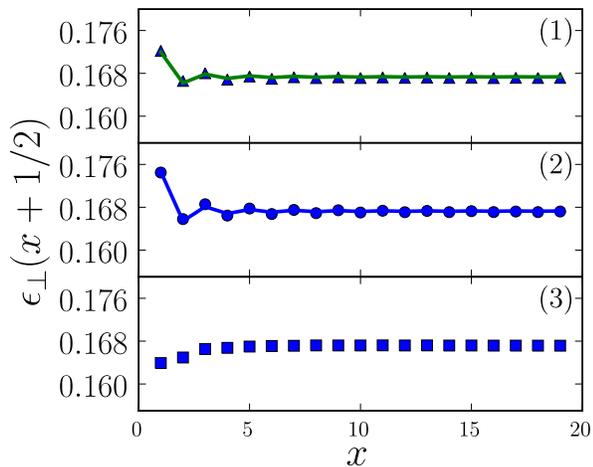}
\caption{(Color online) Mean value of the hopping matrix element on the uncorrelated 
wavefunction versus the distance from the leftmost surface 
layer in geometry (a) with $U_{\rm surface}=20t$ and $U_{\rm bulk} = 14.6642t$ (triangles, panel 1) and $U_{\rm bulk} = 15.9712t$ 
(squares, panel 3). The circles in panel 2 show the hopping for the same simulation that was performed for panel 1, 
but as a function of distance from the right surface of the sample, 
where $U=U_{\rm bulk}=14.6642t$. The results of fit are showed by the solid lines. 
From above, the first and second curves are a plot of Eq.~\eqref{func_hop} with $A=0.1673$, 
$w=-0.0046$ and $A=0.1673$, $w=-0.0074$ respectively.}\label{Fig_surface_hoppings}
\end{figure}

In the previous sections we have derived a simple model to extract the behavior of $Z(z)$ assuming uniform values for 
the hopping matrix elements on the uncorrelated Slater determinant.
Of course the hopping is not uniform, its variation being described in most cases by some Friedel oscillations around the bulk 
value (thin solid lines in Figs.~\ref{Fig_surf_1}-\ref{Fig_barrier_3}).
The Friedel's oscillations arise as a consequence of broken translational symmetry in a Fermi gas, 
i.e. around a single impurity or near an interface.  
An impurity embedded in an electron gas of dimensionality $D$ induces 
oscillations that decay as a power law $1/r^D$ and whose wavevector is twice the Fermi wavevector~\cite{GiulianiVignale}.
The Friedel's oscillations in a $D=3$ electron gas with an interface can be obtained as a  
superposition of Friedel oscillations for a layer of impurities, and one can readily find that, moving perpendicularly 
to the interface over a length $x$, they behave   
at leading order as  
\begin{equation}\label{oscillation}
\frac{\cos{2k_{\rm F}x}}{(2k_{\rm F}x)^2}\,,
\end{equation}
results which is strictly valid for a spherical Fermi surface, although the decay exponent is independent of the 
shape of the Fermi surface.

If we include electron-electron interaction via the Hubbard $U$ and treat it by the Gutzwiller approximation, 
we expect that the Friedel's oscillation will be affected also by the layer-dependence of the quasiparticle weight $Z(z)$. 
Our results show that the faster the change of $Z(z)$, the larger the oscillations. This means that a system with geometry (a) 
and $U_{\rm bulk} \lessapprox U_{\rm crit}$ displays much smoother oscillations that a system with $U_{\rm bulk}\ll U_{\rm crit}$, 
since the spatial dependence of $Z(z)$ is sharper when the bulk interaction parameter is far from criticality.

\begin{figure}[t!]
\includegraphics[width=8.7cm]{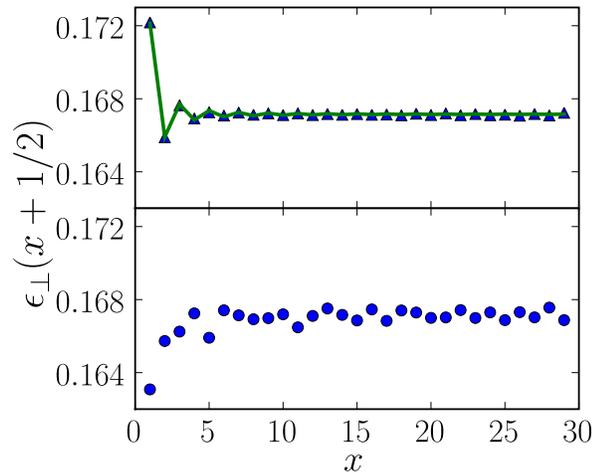}
\caption{(Color online) Plot of the hopping matrix element for a system with 
geometry (b), $U_{\rm left} = 2t$ and $U_{\rm right} = 15.97118t$. 
Upper panel: $x$ is the distance from the junction on the weakly correlated metallic (left) side; lower panel: 
the same on the strongly correlated metallic (right) side. The Friedel oscillations on the weakly correlated side are fitted 
by Eq.~\eqref{func_hop} with $A=0.16715$, $w=0.0050$. On the strongly correlated side the fit was not possible 
for the reasons explained in the text.}\label{Fig_wcmscm_hoppings}
\end{figure}

In light of the spatial dependence of the oscillations predicted by Eq.~\eqref{oscillation}, we fitted our data for the hopping 
$\epsilon_{\perp}(x+1/2)$ perpendicular to the interface and 
in geometries (a) and (b) (see Fig.~\ref{Fig_geometries}) with the function
\begin{equation}\label{func_hop}
A+w\frac{\cos{\pi x}}{x^2}\,,
\end{equation}
where $x$ is the distance from either the surface layer (geometry (a)) or the layer across which $U(z)$ changes 
stepwise (geometry (b)).
The function~\eqref{func_hop} fits the data showed in Figs.~\ref{Fig_surface_hoppings} for a weakly correlated system 
with strongly correlated surface. If the bulk value of $U$ is increased towards $U_{\rm crit}$, the correlation length $\xi$ 
becomes so big that it is hard to identify unambiguously any Friedel's oscillation, as shown in Fig.~\ref{Fig_surface_hoppings} 
panel (3). The function \eqref{func_hop} fits also the data for the hopping on the weakly correlated side of the junction 
in geometry (b), see Fig.~\ref{Fig_wcmscm_hoppings}). 
On the strongly correlated (right) side of the same junction again the correlation length $\xi$ is 
too large and we were not able to make any fit.
 
In conclusion, the inhomogeneity of the interaction parameter $U$ affects the spatial dependence not only of the quasiparticle weight, 
but also of the hopping matrix element on the uncorrelated Slater determinant. The latter displays Friedel's oscillations that rise 
from the breaking of discrete translational symmetry. In any case, when the system is in the close vicinity of the Mott transition, 
the effects of these oscillations are smoothed out as a result of the diverging characteristic length $\xi$ of 
the local quasiparticle weight.


%
\end{document}